\documentclass[aps,prl,preprint,nopacs,superscriptaddress]{revtex4}

\usepackage{graphicx}
\usepackage{verbatim}
\usepackage{mathrsfs}

\usepackage{amsmath,amsfonts,amssymb}
\usepackage{graphicx}

\def\3{2.8in}    
\def\2{2.5in}
\def\4{3.0in}

\def \beq {\begin{equation}}
\def \eeq {\end{equation}}

\begin{document}

\title{Signatures of the Adler-Bell-Jackiw chiral anomaly in a Weyl Fermion semimetal}

\author{Cheng-Long Zhang\footnote{These authors contributed equally to this work.}}\affiliation{International Center for Quantum Materials, School of Physics, Peking University, China}
\author{Su-Yang Xu$^*$}\affiliation {Laboratory for Topological Quantum Matter and Spectroscopy (B7), Department of Physics, Princeton University, Princeton, New Jersey 08544, USA}
\author{Ilya Belopolski$^*$}\affiliation {Laboratory for Topological Quantum Matter and Spectroscopy (B7), Department of Physics, Princeton University, Princeton, New Jersey 08544, USA}

\author{Zhujun Yuan$^*$}\affiliation{International Center for Quantum Materials, School of Physics, Peking University, China}
\author{Ziquan Lin}\affiliation {Wuhan National High Magnetic Field Center, Huazhong University of Science and Technology, Wuhan 430074, China}
\author{Bingbing Tong}\affiliation{International Center for Quantum Materials, School of Physics, Peking University, China}
\author{Guang Bian}\affiliation {Laboratory for Topological Quantum Matter and Spectroscopy (B7), Department of Physics, Princeton University, Princeton, New Jersey 08544, USA}

\author{Nasser Alidoust}\affiliation {Laboratory for Topological Quantum Matter and Spectroscopy (B7), Department of Physics, Princeton University, Princeton, New Jersey 08544, USA}
\author{Chi-Cheng Lee}
\affiliation{Centre for Advanced 2D Materials and Graphene Research
Centre, National University of Singapore, Singapore 117546}
\affiliation{Department of Physics, National University of Singapore,
Singapore 117542}

\author{Shin-Ming Huang}
\affiliation{Centre for Advanced 2D Materials and Graphene Research
Centre, National University of Singapore, Singapore 117546}
\affiliation{Department of Physics, National University of Singapore,
Singapore 117542}

\author{Tay-Rong Chang}
\affiliation{Department of Physics, National Tsing Hua University, Hsinchu 30013, Taiwan}
\affiliation {Laboratory for Topological Quantum Matter and Spectroscopy (B7), Department of Physics, Princeton University, Princeton, New Jersey 08544, USA}
\author{Guoqing Chang}
\affiliation{Centre for Advanced 2D Materials and Graphene Research
Centre, National University of Singapore, Singapore 117546}
\affiliation{Department of Physics, National University of Singapore,
Singapore 117542}

\author{Chuang-Han Hsu}
\affiliation{Centre for Advanced 2D Materials and Graphene Research
Centre, National University of Singapore, Singapore 117546}
\affiliation{Department of Physics, National University of Singapore,
Singapore 117542}

\author{Horng-Tay Jeng}
\affiliation{Department of Physics, National Tsing Hua University, Hsinchu 30013, Taiwan}
\affiliation{Institute of Physics, Academia Sinica, Taipei 11529, Taiwan}

\author{Madhab Neupane}\affiliation {Laboratory for Topological Quantum Matter and Spectroscopy (B7), Department of Physics, Princeton University, Princeton, New Jersey 08544, USA}
\affiliation {Condensed Matter and Magnet Science Group, Los Alamos National Laboratory, Los Alamos, New Mexico 87545, USA}
\affiliation {Department of Physics, University of Central Florida, Orlando, Florida 32816, USA}
\author{Daniel S. Sanchez}\affiliation {Laboratory for Topological Quantum Matter and Spectroscopy (B7), Department of Physics, Princeton University, Princeton, New Jersey 08544, USA}
\author{Hao Zheng}\affiliation {Laboratory for Topological Quantum Matter and Spectroscopy (B7), Department of Physics, Princeton University, Princeton, New Jersey 08544, USA}
\author{Junfeng Wang}\affiliation {Wuhan National High Magnetic Field Center, Huazhong University of Science and Technology, Wuhan 430074, China}
\author{Hsin Lin}
\affiliation{Centre for Advanced 2D Materials and Graphene Research
Centre, National University of Singapore, Singapore 117546}
\affiliation{Department of Physics, National University of Singapore,
Singapore 117542}

\author{Chi Zhang}
\affiliation{International Center for Quantum Materials, School of Physics, Peking University, China}\affiliation{Collaborative Innovation Center of Quantum Matter, Beijing,100871, China}
\author{Hai-Zhou Lu}\affiliation{Department of Physics, South University of Science and Technology of China, Shenzhen, China}
\author{Shun-Qing Shen}\affiliation{Department of Physics, The University of Hong Kong, Pokfulam Road, Hong Kong, China}
\author{Titus Neupert}\affiliation {Princeton Center for Theoretical  Science, Princeton University, Princeton, New Jersey 08544, USA}

\author{M. Zahid Hasan\footnote{Corresponding authors (emails): mzhasan@princeton.edu and gwljiashuang@pku.edu.cn}}\affiliation {Laboratory for Topological Quantum Matter and Spectroscopy (B7), Department of Physics, Princeton University, Princeton, New Jersey 08544, USA}

\author{Shuang Jia$^\dag$}
\affiliation{International Center for Quantum Materials, School of Physics, Peking University, China}\affiliation{Collaborative Innovation Center of Quantum Matter, Beijing,100871, China}

\begin{abstract}
Weyl semimetals provide the realization of Weyl fermions in solid state physics. Among all the physical phenomena that are enabled by Weyl semimetals, the chiral anomaly is the most unusual one. We report signatures of the chiral anomaly in the magneto-transport measurements on the first Weyl semimetal TaAs. We show negative magneto-resistance under parallel electric and magnetic fields, that is, unlike most metals whose resistivity increases under an external magnetic field, we observe that our high mobility TaAs samples become more conductive as a magnetic field is applied along the direction of the current for certain ranges of the field strength. We present unprecedented systematically detailed data and careful analyses, which allow us to exclude other possible origins of the observed negative magneto-resistance. Our transport data, corroborated by photoemission measurements, first-principles calculations, and theoretical analyses, collectively demonstrate signatures of the chiral anomaly in the magneto-transport of TaAs.
\end{abstract}

\date{\today}

\maketitle

\subsection{\large Introduction}

The principles of physics rest crucially on symmetries and their associated conservation laws. Over the past century, physicists have repeatedly observed the violations of apparent conservation laws in particle physics, each time leading to new insights and a refinement of our understanding of nature. One of the most interesting phenomena of this type is the breaking of a conservation law of classical physics by quantum mechanical effects, a so-called anomaly in quantum field theory \cite{Anomaly}. Perhaps the most primitive example is the so-called chiral anomaly associated with Weyl fermions \cite{Weyl, Nielsen1983, ABJ1, ABJ2, Volovik2003}. A Weyl fermion is a massless fermion that carries a definite chirality. Due to the chiral anomaly, the chiral charge of Weyl fermions is not conserved by the full quantum mechanical theory. Historically, the chiral anomaly was crucial in understanding a number of important aspects of the standard model of particle physics. The most well-known case is the triangle anomaly associated with the decay of the neutral pion $\pi^{0}$ \cite{ABJ1, ABJ2}. Despite having been discovered more than 40 years ago, it remained solely in the realm of high-energy physics.

Recently, there has been considerable progress in understanding the correspondence between high energy and condensed matter physics, which has led to deeper knowledge of important topics in physics such as spontaneous symmetry breaking, phase transitions and renormalization. Such knowledge has, in turn, greatly helped physicists and materials scientists to better understand magnets, superconductors and other novel materials, leading to important practical device applications. Here, we present the signatures of the chiral anomaly in a low energy condensed matter Weyl system. In order to measure the chiral anomaly in a solid state system, one needs to find a perturbation that couples differently to the two Weyl fermions of opposite chiralities. This is most naturally realized in a Weyl semimetal, in which the two Weyl cones are separated in momentum space. Recent theoretical and experimental advances have shown that Weyl fermions can arise in the bulk of certain novel semimetals with nontrivial topology \cite{Balents_viewpoint, Wan2011, Murakami2007, TI_book_2014, Ashvin_Review, Hasan_TaAs, Huang2015, Weng2015, MIT_Weyl, TaAs_Ding}. A Weyl semimetal is a bulk crystal whose low energy excitations satisfy the Weyl equation. Therefore, the conduction and valence bands touch at discrete points, the Weyl nodes, with a linear dispersion relation in all three momentum space directions moving away from the Weyl node. The nontrivial topological nature of a Weyl semimetal guarantees that Weyl fermions with opposite chiralities are separated in momentum space (Fig.~\ref{Fig1}\textbf{a}), and host a monopole and an antimonopole of Berry flux in momentum space, respectively (Fig.~\ref{Fig1}\textbf{b}). In this situation, parallel magnetic and electric fields can pump electrons  between Weyl cones of opposite chirality that are separated in momentum space (Fig.~\ref{Fig1}\textbf{a}). This process violates the conservation of the the chiral charge, meaning that the number of particle of left and right chirality are not separately conserved \cite{Nielsen1983, Duval06, Fukushima, Chiral_Burkov, Aji, Zynzin, Grushin, Qi_review, Jho13, Son13, Burkov14}, giving rise to an analog of the chiral anomaly in a condensed matter system. Apart from this elegant analogy and correspondence between condensed matter and high energy physics, the chiral anomaly also serves as a crucial transport signature for Weyl fermions in a Weyl semimetal phase. Furthermore, theoretical studies have recently suggested that it has potential applications \cite{Nonlocal}.

In this paper, we perform magneto-transport experiments on the first Weyl semimetal TaAs \cite{Hasan_TaAs, Huang2015, Weng2015, TaAs_Ding}. We observe a negative longitudinal magnetoresistance (LMR) in the presence of parallel magnetic and electric fields, which is indicative of the chiral anomaly due to Weyl fermions. On the other hand, due to the complicated nature of the magnetoresistence \cite{GMR, Pippard, Ag2Se, Argyres, Kikugawa, Das_Sarma, Haizhou, InSb, Burkov_2015, Maslov, Spivak2015}, an unambiguous demonstration of the chiral anomaly remains lacking despite the volume of works reporting negative LMR \cite{Korean_BiSb, IOP_Chiral, Shuang_Chiral, Ong_Chiral, CdAs_Chiral, Yan_Chiral}. Our unprecedented data and careful analyses, which go beyond a simple observation of a negative LMR, allow us to systematically exclude other possible origins for the observed negative LMR. These data strongly support the chiral anomaly due to Weyl fermions in TaAs. Our studies demonstrate a low-energy platform where the fundamental physics of Weyl fermions and quantum anomalies now can be studied in a piece of solid-metal \cite{Duval06, Fukushima, Chiral_Burkov, Aji, Zynzin, Grushin, Qi_review, Jho13, Son13, Burkov14, Nonlocal}.

\subsection{\large Results}

\bigskip
\textbf{ARPES band structure}

\begin{figure*}
\centering
\includegraphics[width=17cm]{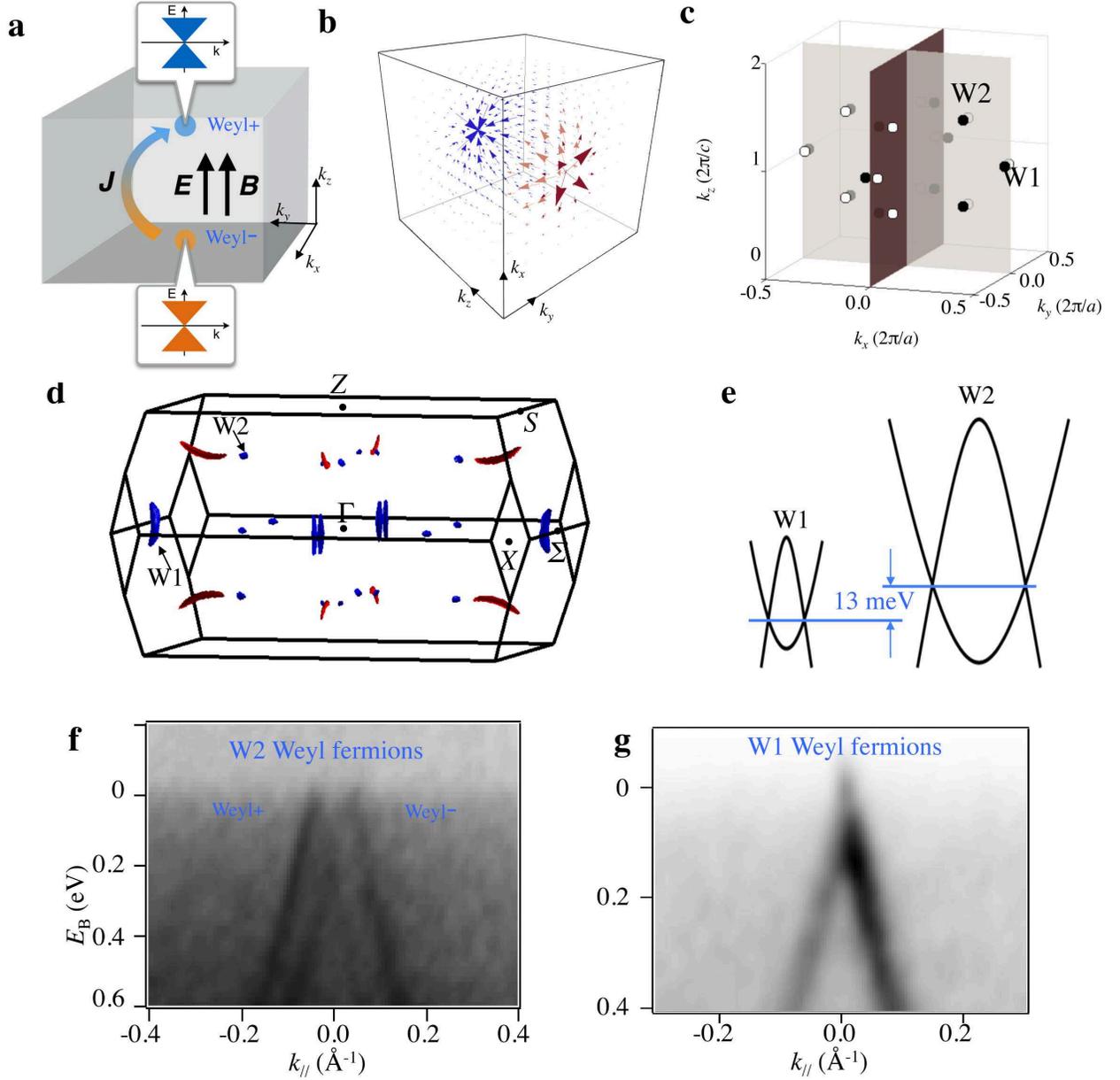}
\caption{\label{Fig1}\textbf{Electronic band structure of the Weyl semimetal TaAs.} \textbf{a,} Schematics of the separation of the pairs of Weyl fermions in a Weyl semimetal with opposite chiralities in momentum space, which is a direct consequence of its nontrivial topological nature. \textbf{b,} Distribution of the Berry curvature near two Weyl nodes in momentum space with the opposite chiralities. \textbf{c,} The location of the Weyl nodes in the first Brillouin zone. \textbf{d,} First principles calculated constant energy contour of TaAs. The energy is set at about 5 meV above the energy of the W2 Weyl nodes. \textbf{e,} Calculated energy dispersions of the W1 and the W2 Weyl cones. \textbf{f,} ARPES measured energy dispersions of the W2 Weyl cones. \textbf{g,} ARPES measured energy dispersions of the W1 Weyl cones.}
\end{figure*}

We start by presenting the key aspects of the bulk band structure of TaAs both in theory and in experiment. According to our first-principles calculation \cite{Huang2015, Weng2015}, in total there are 24 bulk Weyl cones. We denote the 8 Weyl nodes that are located on the $k_z=\frac{2\pi}{c}$ as W1 and the other 16 nodes that are away from this plane as W2 (Fig.~\ref{Fig1}\textbf{c}). There is a 13 meV offset between the energies of the W1 and W2 Weyl nodes (Fig.~\ref{Fig1}\textbf{e}). The pockets that arise from the Weyl fermions are shown in blue in Fig.~\ref{Fig1}\textbf{d}. Apart from the Weyl cones, there are additional (non-Weyl) hole-like bands crossing the Fermi level shown by the red ring-shaped contours in Fig.~\ref{Fig1}\textbf{d}.

We independently study the bulk electronic structure via angle-resolved photoemission spectroscopy (ARPES). This is important because relying entirely on numerical band structure calculations is not conclusive. Particularly, numerical band calculations have little power in predicting the position of the chemical potential of real samples, which is crucial for transport experiments. Figure~\ref{Fig1}\textbf{f} shows an $E-k_{||}$ dispersion map that cuts across the two nearby W2 Weyl cones. The dispersion map reveals two linearly dispersive bands. The $k$-space distance between the two crossing points is about $0.08$ $\textrm{\AA}^{-1}$, consistent with the calculated results. More importantly, our ARPES measurement shows that the native chemical potential of the samples is very close to the energy of the W2 Weyl nodes. Our data also reveals the W1 Weyl cones. As shown in Fig.~\ref{Fig1}\textbf{g}, the energy of the W1 Weyl node is below that of the W2 Weyl node (Fig.~\ref{Fig1}\textbf{f}), consistent with band calculation results. Systematic ARPES data can be found in the Supplementary Figure 1 and Supplementary Note 1. We also observe the trivial hole bands in ARPES. The essential observations are listed as follows: (1) There are three types of bands at the Fermi level, the W1 and W2 Weyl nodes and a trivial hole like band. (2) The native chemical potential is close to the energy of the W2 Weyl nodes, which is 13 meV higher than that of the W1 Weyl nodes. (3) Therefore, the W1 Weyl cones form electron-like pockets, the trivial hole-like bands form hole-like pockets, and the W2 Weyl cones have low carrier concentration, which can be electron- or hole-like depending on the specific position of the native chemical potential with respect to the W2 Weyl node in each sample batch.

\bigskip
\textbf{Quantum oscillation data}

\begin{figure*}
\centering
\includegraphics[width=16cm]{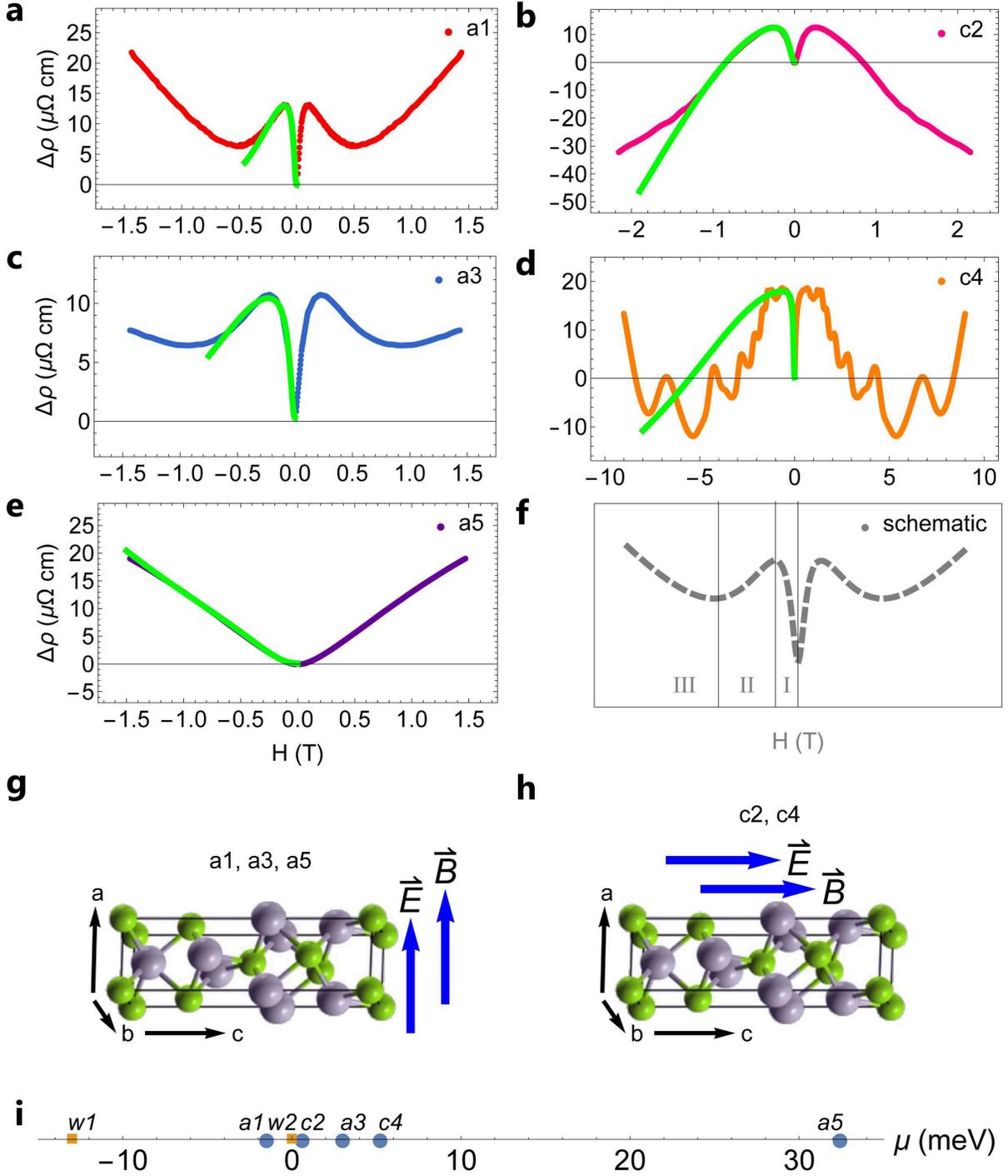}
\caption{\label{Fig2}\textbf{Observation of negative longitudinal magneto-resistances.} \textbf{a-e,} LMR data at $T=2$ K for samples a1, c2, a3, c4, and a5, respectively. The green curves are the fits to the LMR data in the semiclassical regime. The $y$ axes of panels (a-e) are the change of the resistivity with respect to the zero-field resistivity, $\Delta\rho=\rho(B)-\rho(B=0)$.}
\end{figure*}

We have performed magneto-transport measurements on our TaAs samples, in order to probe the band structure at the Fermi level (Supplementary Figure 2 and Supplementary Note 2). Our Hall data indeed reveal a coexistence of electron and hole carriers. We obtained critical band parameters, such as Fermi wavevector, Fermi velocity, chemical potential, carrier mobilities, etc., from the Shubnikov de-Haas (SdH) oscillation data. All band parameters obtained from SdH oscillation are consistent with first-principles calculation and ARPES results. Most importantly, this enables us to determine the position of the chemical potential with respect to the Weyl nodes, as shown in Figures~\ref{Fig2}\textbf{i}. We name the samples by a letter (a or c) followed by a number (1-5). The letter a or c means that the electrical current is along the crystallographic $a$ or $c$ axes. The number 1-5 refers to a sequence of the samples' chemical potential with respect to the energy of the W2 Weyl node from below the node to above the node (Fig.~\ref{Fig2}\textbf{i}). We also provide the Fermi energy of the TaAs samples determined by different approaches in Supplementary Note 3 and Supplementary Table 1.

\bigskip
\textbf{Longitudinal magnetoresistance}

We now present our LMR data, without a pre-biased assumption of their origin. Figs.~\ref{Fig2}\textbf{a-e} show the LMR data on 5 different batches of samples. The LMR data show three main features as a function of the magnetic field, as schematically drawn in Fig.~\ref{Fig2}\textbf{f}. At very small fields close to $B=0$, we observe a sharp increase of the LMR. Following the sharp increase, the LMR is found to decrease in an intermediate $B$ field range. This is the negative LMR. While further increasing the $B$ field, the LMR starts to increase again. We note that because features I and II likely have independent origins, the LMR is not necessarily absolutely negative (we do, however, observe absolute negative LMR in samples c2 and c4). Hence, most precisely, the negative LMR means a negative response of the LMR as one increases the $B$ field. In addition to these general features, we observe other more sample dependent features: For sample c4, our data show clear quantum oscillations at a quite wide $B$ field range of $0.5$ T $\leq{B}\leq8$ T. For other samples, the quantum oscillations are much weaker but they are still visible. For sample a5, the LMR increases monotonically as a function of the $B$ field. No negative LMR is observed.

We study the systematic dependence of the LMR on different parameters, including temperature, the angle between the $\vec{E}$ and $\vec{B}$ fields, and the direction of the current with respect to the crystallographic axis. The temperature dependent data are shown in Fig.~\ref{Fig3}\textbf{a} for sample a1. Most notably, the negative LMR (feature II) shows a strong temperature dependence. At higher temperatures, e.g. $T\geq50$ K, the negative LMR vanishes. The dependence on the angle between the $\vec{E}$ and $\vec{B}$ fields are shown in Figs.~\ref{Fig3}\textbf{b-e}. Our data show that the negative MR exhibits a very strong angular dependence. It becomes quickly suppressed as one varies the direction of the magnetic $\vec{B}$ field away from that of the electric $\vec{E}$ field. The dependence on the direction of the current with respect to the crystallographic direction is presented in Fig.~\ref{Fig2}. The measurements were performed with current along the crystallographic $a$ axis for samples a1, a3 and a5, and with current along the $c$ axis for samples c2 and c4. In both cases, the negative LMR is observed except for sample a5, whose chemical potential is far away from the energy of the Weyl nodes (Fig.~\ref{Fig2}\textbf{i}).

\bigskip
\textbf{Origins of the negative longitudinal magnetoresistance}

We now use these observations to understand the origin of the negative LMR. First, it is well-known that a negative LMR can arise in magnetic materials \cite{GMR}. This obviously does not apply to our non-magnetic TaAs samples. The second possible origin is more classical due to geometry or size effects of the samples, such as the current jetting effect \cite{Pippard, Ag2Se}. These geometrical MR effects are also not consistent with our data, because they do not vanish quickly as one raises temperature \cite{Ag2Se}, and furthermore we have carefully shaped our samples to exclude the geometrical effects (Supplementary Figure 3 and Supplementary Note 4). Third, we observe the negative LMR with current flowing both along the crystallographic $a$ and $c$ axes. We note that TaAs has a tetragonal lattice. Hence the $a$ and $c$ axes represent the largest anisotropy that the system could offer. The fact that the negative LMR is observed along both $a$ and $c$ axes proves that anisotropies in the system cannot explain our data \cite{Kikugawa}. Fourth, in the quantum limit, negative LMR can arise from the chiral, quasi one-dimensional character of the Landau levels that are formed by the band structure under magnetic fields. Essentially, it was predicted \cite{Argyres, Das_Sarma} that a negative LMR can arise in any 3D metal irrespective of its band structure if the sample is in the ultra quantum limit, which means that one has $\omega\tau\gg1$ ($\omega$ is the cyclotron frequency and $\tau$ is the transport lifetime) and that the chemical potential only crosses the lowest Landau level (LLL) (the Landau level index $N=0$, see Supplementary Figure 4). This has been observed in doped semiconductor samples \cite{InSb}. We have carefully checked whether our negative LMR is due to this mechanism. Particularly, one needs to be careful about the trivial hole-like bands in TaAs, because if they were in the ultra quantum limit then it would have been entirely possible that the observed negative LMR were due to these trivial bands, rather than due to Weyl fermions in our samples. We note that the negative LMR are observed at small magnetic fields (e.g. $0.1$ T $\leq{B}\leq0.5$ T for sample a1). We have checked the $\omega\tau$ and the Landau level index $N$ of our samples quantitatively (Supplementary Table 2), and our results show that all samples are always in the semiclassical limit at the small magnetic fields where the negative LMR are observed. Therefore, our data are inconsistent with this origin \cite{Argyres, Das_Sarma}. Fifth, a recent theoretical work has predicted a linear B dependent magneto-conductivity in small fields \cite{Haizhou3}. However, this is also inconsistent with our data because predicted linear B dependent magneto-conductivity requires the system to lie in the ultra-quantum limit. That is, only the lowest Landau band crosses the Fermi level, which is clearly not the case for our systems under study. Finally, in the semiclassical limit, nonzero LMR can arise from finite Berry curvature, as follows from the semiclassical equations of motion. Having excluded all other possibilities, we are led to conclude that our observed negative LMR must have this origin. However, Ref. \cite{Burkov_2015} showed that any Berry curvature not associated with Berry monopoles, i.e., not coming from Weyl nodes, will contribute a positive LMR. The observed negative LMR should thus be attributed to Weyl nodes, in accordance with the theoretical analysis of Refs. \cite{Son13, Burkov14, Burkov_2015}. To confirm this picture independent of the assumption of effective low-energy Weyl Hamiltonians, we have studied the contribution of Berry curvature from each band carefully in a first-principles derived model for TaAs (see Figures.~\ref{Fig4}\textbf{c,d} and Supplementary Figure 5 and Supplementary Note 5.). Our results show that in our TaAs system the Berry curvature almost entirely arises from the Weyl cones.

\bigskip
\textbf{The chiral anomaly}

With such a conclusion, we are entitled to fit our LMR data with a semiclassical magnetoconductance formula that includes the contribution from Weyl nodes due to their Berry curvature. Specifically, we use the following equation.
\begin{equation}
\sigma_{xx}(B)=8{C_{\textrm{W}}}B^2-C_{\textrm{WAL}}\bigg(\sqrt{B}\frac{B^2}{B^2+B_{\textrm{c}}^2}+\gamma{B}^2\frac{B_{\textrm{c}}^2}{B^2+B_{\textrm{c}}^2}\bigg)+\sigma_0
\end{equation}
All coefficients are positive. The first term $\sigma^{\textrm{chiral}}={C_{\textrm{W}}}B^2$ is due to the Weyl fermions and will lead to a $B^2$ dependent negative LMR. This term was systematically studied by transport theories in Ref. \cite{Son13, Burkov14}. The chiral coefficient is $C_{\textrm{W}}=\frac{e^4\tau_{\textrm{a}}}{4\pi^4\hbar^4g(E_{\textrm{F}})}$ \cite{Son13, Burkov14}, where $g(E_{\textrm{F}})$ is the density of states at the Fermi level, $\tau_{\textrm{a}}$ is the axial charge relaxation time, and the additional factor of 8 is because we have 8 pairs of W2 Weyl nodes. All remaining terms contribute to positive LMR in the semiclassical regime. The $C_{\textrm{WAL}}$ term arises from the 3D weak antilocalization (WAL) effect of the Weyl cones, which accounts for the initial steep uprise of the LMR at small magnetic fields. The 3D WAL is known to have a $-B^2$ dependence near zero field and $-\sqrt{B}$ dependence at higher fields \cite{Haizhou2}. So we include a critical field $B_{\textrm{c}}$ that characterizes a crossover. For the four samples a1, c2, a3, c4, the increase of the LMR at small magnetic fields are 230\%, 5\%, 156\%, and 47\% comparing to the zero field resistance. Particularly, the increase for samples a1 and a3 is larger than 100\%, which is usually not expected from the WAL scenario. On the other hand, we do notice that the increase is quite sample dependent, and that a similarly large increase ($\sim100\%$) of the magneto-resistance has also been reported in a concurrent transport work on TaAs \cite{IOP_Chiral}. In this work, we fit this initial uprise of the LMR by the WAL effects, but the anomalously large increase in samples a1, a3 and also in Ref. \cite{IOP_Chiral} remain an theoretically open question that needs further investigation, which does not affect our main conclusion, i.e., signatures of the chiral anomaly. Finally, the  $\sigma_0$ term is the positive LMR that arises from the Drude conductivity of conventional charge carriers present in TaAs. In parallel fields, the Lorentz force is zero so the Drude conductivity is a constant. More systematic details regarding the fitting are presented in Supplementary Figure 6 and Supplementary Note 6.

The fitting results are shown by the green curves in Figs.~\ref{Fig2}\textbf{a-e} and Figs.~\ref{Fig3}\textbf{a,b}. It can be seen that the fitting works well for the small $B$ field region which includes the negative LMR. This is reasonable because the fitting formula is derived in the semiclassical limit. The angle dependence of the chiral coefficient $C_{\textrm{W}}$ is shown in Fig.~\ref{Fig4}\textbf{b} for sample a3, which demonstrates $C_{\textrm{W}}$ is only significant in the presence of parallel electric and magnetic fields. More importantly, we study the chemical potential dependence of the LMR data. Our fitting captures quantitatively the relative size of the low-field positive LMR and the higher-field negative LMR as a function of chemical potential. We plot this ratio as a dimensionless quantity in Fig.~\ref{Fig4}\textbf{a}. We find that despite the simple form of the fitting formula, the different measurement geometries for the different samples, the presence of large quantum oscillations in sample c4 and large differences in the absolute resistivities of different samples, the chiral anomaly ratio scales as $1/E_{\textrm{F}}^2$. It is remarkable that this fitting result matches the simplest theoretical model for a Weyl point, where the Berry curvature $\Omega \propto 1/E_{\textrm{F}}^2$. We emphasize that this provides powerful evidence that the negative LMR is due to the Weyl fermions. Note that the specific expression of the chiral coefficient, $C_{\textrm{W}}=\frac{e^4\tau_{\textrm{a}}}{4\pi^4\hbar^4g(E_{\textrm{F}})}\propto\frac{1}{E_{\textrm{F}}^2}$, is a result of the linear dispersion and the specific Berry curvature distribution of the Weyl cones (see Fig.~\ref{Fig1}\textbf{b}). Especially, the energy of the W2 Weyl nodes is nothing special for the trivial hole bands. Thus if the negative LMR arose from the hole bands, then the chiral coefficient $C_{\textrm{W}}$ would have not increased dramatically as the chemical potential approaches the energy of the W2 Weyl nodes (Fig.~\ref{Fig4}\textbf{a}). Therefore, the obtained $\frac{1}{E_{\textrm{F}}^2}$ dependence of the chiral coefficient (Fig.~\ref{Fig4}\textbf{a}) provides a unique demonstration that our negative LMR is due to the Weyl fermions, because the $\frac{1}{E_{\textrm{F}}^2}$ dependence reveals the details of the band dispersion and Berry curvature distribution of the Weyl cones, not just the fact that the bands have some nonzero Berry curvature. We use the above data and analyses to further exclude the possibility of negative LMR due to weak localization arising from the intervalley scattering. Our data is not due to weak localizations for the following reasons: First, weak localization does not have a strong $E_\textrm{F}$ dependence, let alone the dramatic $\frac{1}{E_{\textrm{F}}^2}$ dependence observed in our data. Second, it has been theoretically shown that the magnetoconductivity without the chiral anomaly is always monotonic, even though the intervalley scattering can induce a negative LMR arising from the weak localization \cite{Haizhou2}. This is not consistent with our data, which means that, without the chiral anomaly, only weak localization/anti-localization cannot explain our data.

Up to here, we have demonstrated that the negative LMR arises from the nonzero Berry curvature of the Weyl fermions in our TaAs samples. We now establish the connection between our data and the chiral anomaly, the nonconservation of the electron quasiparticle number of the Weyl cones with a given chirality. In a real Weyl semimetal sample, this can be understood by two crucial components, the axial charge pumping effect and the axial charge relaxation, as schematically shown in Fig.~\ref{Fig4}\textbf{f}. The charge pumping effect means that a nonzero $\mathbf{E}\cdot \mathbf{B}$ can pump charges from one Weyl cone to the other, leading to an imbalance of the quasiparticle number of the Weyl cones with the opposite chiralities. This effect is well established to occur between Weyl nodes of different chiral charge \cite{Burkov14, Burkov_2015, Son13}, which are monopoles of Berry field strength in momentum space. We have directly show the nontrivial Berry curvature monopoles associated with the Weyl fermions via our LMR transport data. The axial charge pumping creates an out-of-equilibrium quasiparticle distribution between the Weyl cones with opposite chiralities. To form a steady state, it is counteracted by the relaxation of the axial charge disproportionation through scattering between the Weyl nodes. The relaxation is characterized by a time scale, the axial charge relaxation time $\tau_{\textrm{a}}$. From our negative LMR data, we directly obtain the axial charge relaxation time $\tau_{\textrm{a}}$ (Fig.~\ref{Fig4}\textbf{e}). The nonzero axial charge relaxation time $\tau_{\textrm{a}}$ not only directly demonstrates the axial charge relaxation, but also confirms the existence of the axial charge pumping because these two are directly coupled, which means that one cannot exist alone if the other is absent. 

We can directly obtain the axial charge relaxation time $\tau_{\textrm{a}}$, which serves as the critical physical quantity that characterizes the chiral anomaly, from the chiral coefficient $C_{\textrm{W}}$ using the relationship $C_{\textrm{W}}=\frac{e^4\tau_{\textrm{a}}}{4\pi^4\hbar^4g(E_{\textrm{F}})}$ \cite{Son13, Burkov14}. In Fig.~\ref{Fig3}\textbf{a}, we present fitting results as a function of temperature for sample s1. We use the fitting coefficients $C_{\textrm{W}}$ to obtain the axial charge relaxation time as a function of temperature, presented in Fig.~\ref{Fig4}\textbf{e}. We find that the $\tau_{\textrm{a}}$ rapidly decays to zero with increasing temperature. This decay of $\tau_{\textrm{a}}$ corresponds to the decay of the negative LMR with increasing temperature in the raw data and is expected because scattering typically increases with temperature. We obtain an axial charge relaxation time $\tau_{\textrm{a}}=5.96\times10^{-11}$ s for sample a1 at $T=2$ K (Fig.~\ref{Fig4}\textbf{e}). Note that this $\tau_{\textrm{a}}$ is associated with the W2 Weyl cones because the Fermi level is very close to the W2 nodes. On the other hand, it is difficult to obtain the transport life time of the W2 Weyl cones because the density of states at the Fermi level is dominated by contributions from the W1 Weyl cones and the trivial hole bands (Fig.~\ref{Fig1}\textbf{d}). Therefore, we estimate the quasiparticle life time associated with the W2 Weyl cones via $\tau\simeq\hbar/E_{\textrm{F}}=7.04\times10^{-13}$ s for sample a1. We see that the axial charge relaxation time $\tau_{\textrm{a}}$ is much longer than the quasiparticle life time $\tau$. The imbalance of population due to the axial charge pumping can be also estimated by the uncertainty principle $\Delta\mu=\hbar/\tau_{\textrm{a}}\simeq0.011$ meV. At $E_{\textrm{F}}=-1.5$ meV, the density of states per W2 Weyl cone is $g(E_{\textrm{F}})=1.6\times10^{16}$ states/(eV$\cdot$cm$^3$). Therefore, we estimate the chiral charge, the non-conservation of the quasi-particle number of the Weyl cone with a given chirality, to be $\Delta\mu\times{g(E_{\textrm{F}}})=1.6\times10^{14}$. This directly characterizes the chiral anomaly in our Weyl semimetal TaAs sample.

\subsection{\large Discussion}

We emphasize the critical logical sequences that are key to our demonstration. Unlike previous studies, we do not assume that the negative LMR arises from the chiral anomaly \cite{IOP_Chiral, Shuang_Chiral, Ong_Chiral, CdAs_Chiral, Yan_Chiral}. In order to demonstrate the chiral anomaly, it is critically important to consider all possible origins for a negative LMR and to discuss how one can distinguish each of the other origins from the chiral anomaly. This has been achieved for the first time here. We first excluded the geometry and spin(magnetic) effects. Then we show that our observed LMR is not in the quantum (large $B$ field) limit, in which the Fermi energy crosses only the lowest Landau level. This is important because the LMR in the quantum (large $B$ field) limit can be negative or positive depending on specific scenarios, such as the band dispersion and nature of the impurities \cite{Argyres, Das_Sarma, Haizhou}. In fact, it is even theoretically shown that the Weyl cones that respect time-reversal symmetry can contribute a positive (not a negative) LMR in the quantum limit if the field dependence of the scattering time and Fermi velocity of the Landau bands is fully respected \cite{Haizhou}. Therefore, observing a negative LMR in the large-field quantum limit \cite{Yan_Chiral} may not be a compelling signature of Weyl fermions. In the semiclassical (small $B$ field) limit, after excluding the geometry and magnetic effects, one can avoid ambiguities in the physical interpretation since a negative LMR can only arise from a nonzero Berry curvature \cite{Son13, Burkov14, Burkov_2015, Maslov}. In fact, it has been shown that the LMR from a band with zero Berry curvature will always be positive \cite{Maslov}. However, we emphasize that the negative LMR in the semiclassical limit is only a signature of the Berry curvature but it is not unique to Weyl fermions. The systematic (temperature, angular, current direction) dependence cannot distinguish the negative LMR due to Weyl fermions from the negative LMR due to other band structures with nonzero Berry curvature because they are expected to show the same qualitative dependence independent of Weyl or non-Weyl bands. In order to uniquely attribute the negative LMR to Weyl fermions, we discovered here that it is crucial to obtain comprehensive information about the band structure. Specifically, first we have shown that the Berry curvature in our TaAs is dominated by the Weyl cones. Second, the chiral coefficient has a $\frac{1}{E_{\textrm{F}}^2}$ dependence. These two pieces of evidence, together with the full systematics of the datasets uniquely presented here, provides strong signatures of the chiral anomaly of Weyl fermions.

\section{Methods:}
\textbf{Sample growth and electrical transport:} High quality single crystals of TaAs were grown by the standard chemical vapor transport method as described in Ref. \cite{Murray}. TaAs crystals were structurally characterized by powder X-ray diffraction to confirm bulk quality, and to determine (001) crystal face. A small portion of the obtained samples were ground into fine powders for X-ray diffraction measurements on Rigaku MiniFlex 600 with Cu $K_{\alpha}$ ($40$ kV, $15$ mA; $\lambda=0.15405$ nm) at room temperature, and then refined by a Rietica Rietveld program. Magneto-transport measurements were performed using a Quantum Design Physical Property Measurement System. High-field electrical transport measurements were carried out using a pulsed magnet of $50$ msec in Wuhan National High Magnetic Field Center. All the measurements were carried out from -9 T to 9 T or -56 T to 56 T.

\textbf{Angle-resolved photoemission spectroscopy:} The soft X-ray ARPES (SX-ARPES) measurements were performed at the ADRESS Beamline at the Swiss Light Source in the Paul Scherrer Institut in Villigen, Switzerland using photon energies ranging from 300 to 1000 eV \cite{SXARPES}. The sample was cooled down to 12 K to quench the electron-phonon interaction effects reducing the $k$-resolved spectral fraction. The energy and angle resolution was better than 80 meV and $0.07^{\circ}$, respectively. Vacuum ultraviolet ARPES measurements were performed at beamlines 4.0.3, 10.0.1 and 12.0.1 of the Advanced Light Source at the Lawrence Berkeley National Laboratory in Berkeley, California, USA, Beamline 5-4 of the Stanford Synchrotron Radiation Light source at the Stanford Linear Accelerator Center in Palo Alto, California, USA and Beamline I05 of the Diamond Light Source in Didcot, UK, with the photon energy ranging from 15 eV to 100 eV. The energy and momentum resolution was better than 30 meV and 1\% of the surface Brillouin zone. 

\textbf{Theoretical calculations:} First-principles calculations were performed by the OPENMX code within the framework of the generalized gradient approximation of density functional theory \cite{Perdew}. Experimental lattice parameters were used [48], and the details for the computations can be found in our previous work in Ref. \cite{Murray}. A real-space tight-binding Hamiltonian was obtained by constructing symmetry-respecting Wannier functions for the As $p$ and Ta $d$ orbitals without performing the procedure for maximizing localization.

\bigskip
\bigskip
\textbf{Acknowledgements}

M.Z.H., S.-Y.X., and I.B. thank I. Klebanov, A. Polyakov and H. Verlinde for theoretical discussions. T.N. thanks A. G. Grushin for discussions. S.J. thanks J. Xiong and F. Wang for valuable discussions, and C.-.L.Z. and Z.Y. thank Y. Li and J. Feng for using instruments in their groups. The work at Princeton and Princeton-led synchrotron-based measurements were supported by Gordon and Betty Moore Foundation through Grant GBMF4547 (Hasan). S. J. was supported by the National Basic Research Program of China (Grant Nos. 2013CB921901 and 2014CB239302) and by the Opening Project of Wuhan National High Magnetic Field Center (Grant No.PHMFF2015001), Huazhong University of Science and Technology. C. Z. was supported by the National Science Foundation of China (Grant No.11374020). H.L. acknowledges the Singapore National Research Foundation for the support under NRF Award No. NRF-NRFF2013-03. S.-Q.S. was supported by the Research Grant Council, University Grants Committee, Hong Kong under Grant No. 17303714. H.-Z.L. was supported by the Natural Science Foundation of China under Grant No. 11574127. We gratefully acknowledge J. D. Denlinger, S. K. Mo, A. V. Fedorov, M. Hashimoto, M. Hoesch, T. Kim,  and V. N. Strocov for their beamline assistance at the Advanced Light Source, the Stanford Synchrotron Radiation Lightsource, the Diamond Light Source, and the Swiss Light Source. S.M.H., G.C., T.R.C, and H.L.'s visits to Princeton University were funded by the U.S. Department of Energy (DOE), Office of Science, Basic Energy Sciences (BES) under the funding number DE-FG-02-05ER46200.

\bigskip
\bigskip
\textbf{Author contributions}
\newline

C.-L.Z., performed the electrical transport experiments with the help from Z.Y., Z.L., B.T., J.W., C.Z., and S.J.. S.-Y.X., I.B., G.B. conducted the ARPES experiments with the assistance from N.A., M.N., D.S.S., H.Z., and M.Z.H.. C.-L.Z., Z.Y., and S.J. grew the single crystal samples; C.-C.L., S.-M.H., T.-R.C., G.C., C.-H.H., H.-T.J., and H.L. performed first-principles band structure calculations; H.-Z.L., S.-Q.S., and T.N. did theoretical analyses; H.-Z.L. and S.-Q.S. proposed the fitting formula for the weak anti-localization. I.B. performed the fitting to the magneto-resistance data. S.-Y.X., M.Z.H., and S.J. were responsible for the overall direction, planning and integration among different research units.

\section{Additional information}
The authors declare no competing financial interests.

\clearpage

\clearpage

\addtocounter{figure}{-1}
\begin{figure*}[t!]
\caption{\textbf{h,} A schematic drawing of the LMR data to show the three important features (I-III) observed in our data as a function of the magnetic field. \textbf{g,} Measurement geometry for samples a1, a3, and a5. \textbf{h,} Measurement geometry for samples c2 and c4. \textbf{i,} Position of the samples' chemical potential with respect to the energy of the Weyl nodes obtained from SdH oscillation measurements.}
\end{figure*}

\begin{figure*}
\centering
\includegraphics[width=15cm]{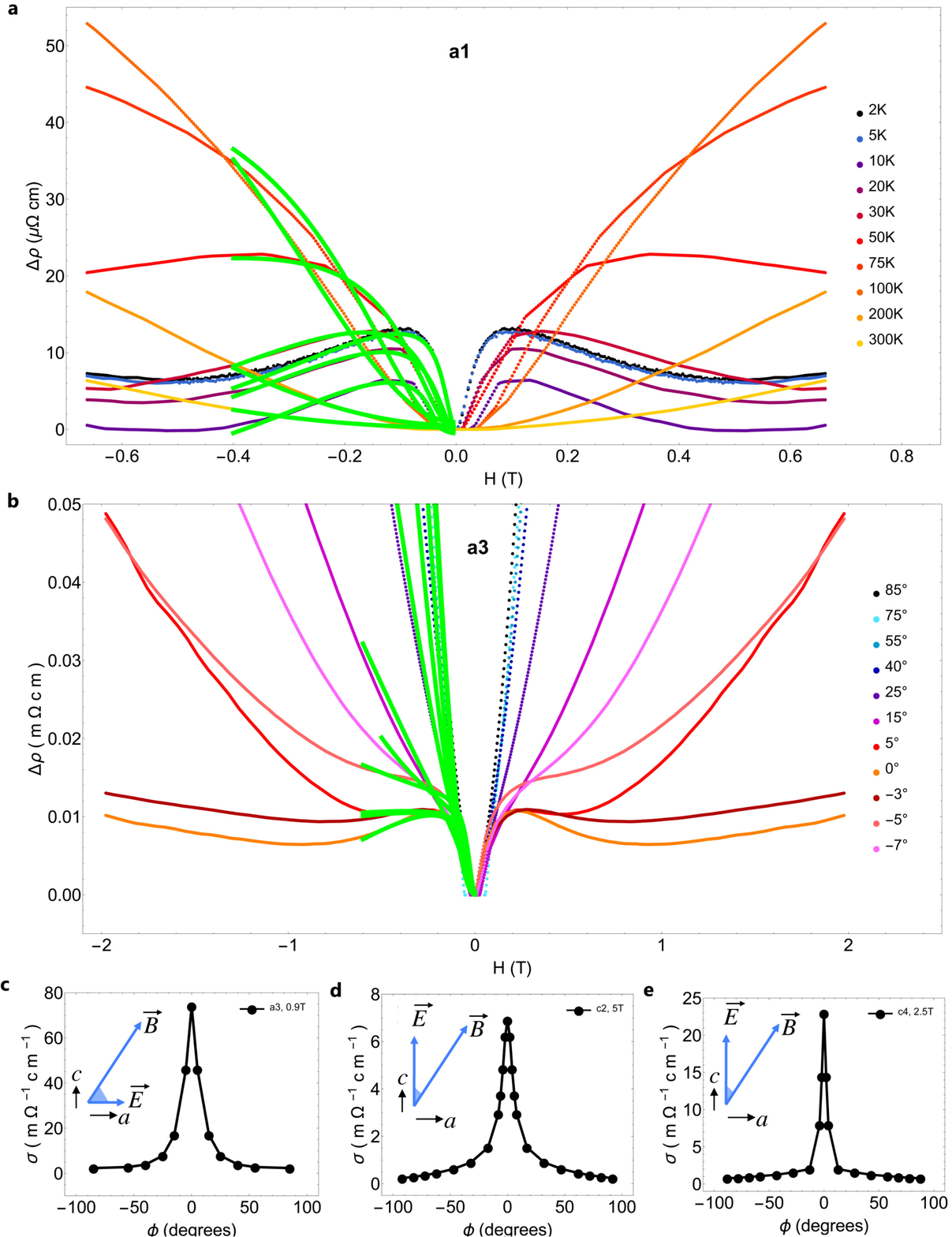}
\caption{\label{Fig3}\textbf{Systematic dependence of the negative longitudinal magneto-resistances.}
\textbf{a,} Temperature dependent LMR data for sample a1. \textbf{b,} Magnetoresistance data as a function of the angle between the $\vec{E}$ and $\vec{B}$ fields. The green curves of panels (a,b) are the fits to the LMR data in the semiclassical regime. The $y$ axes are the change of the resistivity with respect to the zero-field resistivity, $\Delta\rho=\rho(B)-\rho(B=0)$. \textbf{c-e,} The magnetoresistance as a function of the angle for samples c2, a3, and c4 at a fixed field.}
\end{figure*}

\begin{figure*}
\centering
\includegraphics[width=15cm]{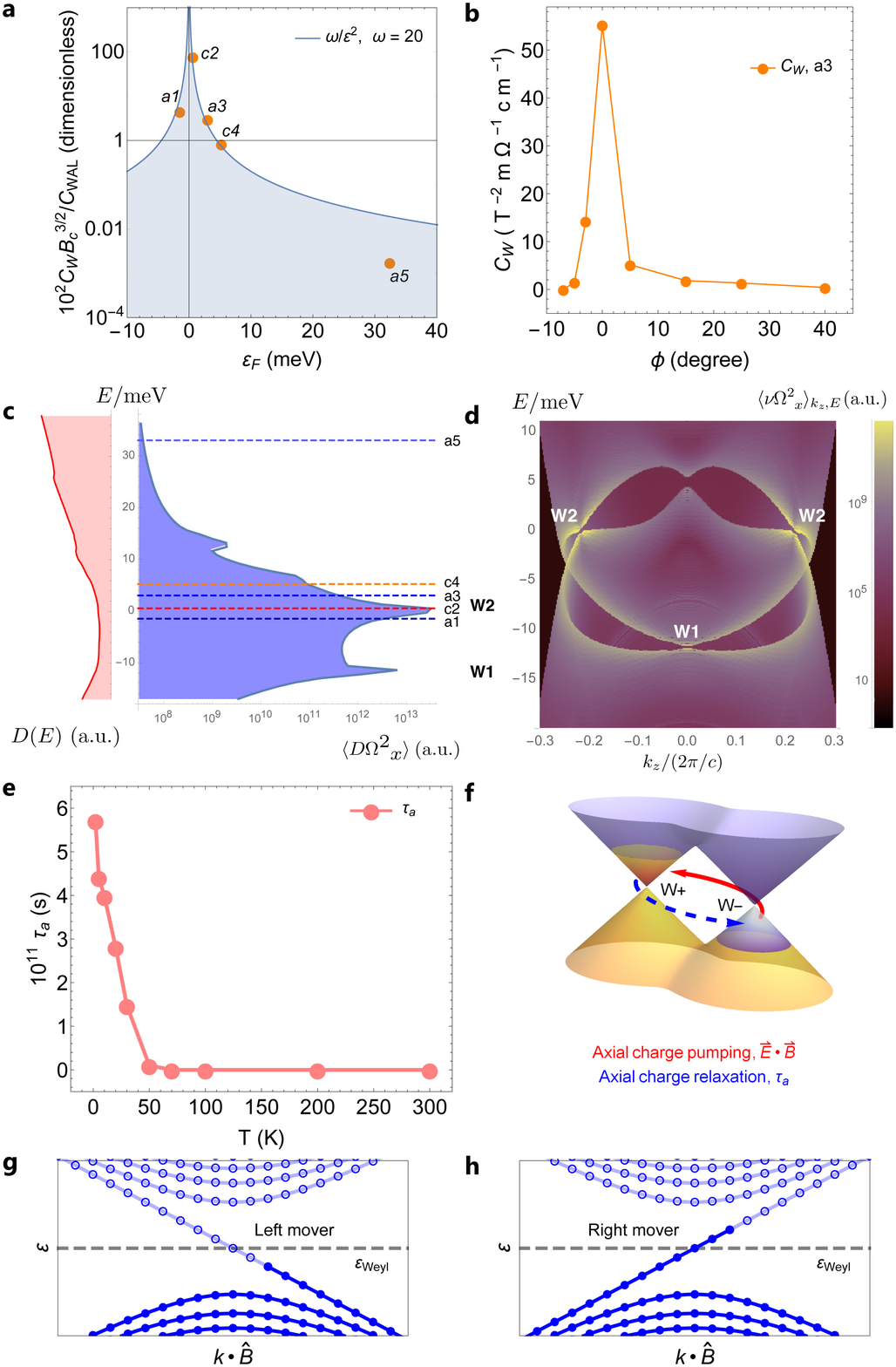}
\caption{\label{Fig4}\textbf{Signatures of the chiral anomaly due to Berry curvature of the Weyl fermions.} }
\end{figure*}
\addtocounter{figure}{-1}
\begin{figure*}[t!]
\caption{\textbf{a,} Chemical potential $E_{\textrm{F}}$ dependence of the chiral coefficient $C_{\textrm{W}}$. We expect the chiral coefficient $C_{\textrm{W}}$ to decay as a function of $\frac{1}{E_{\textrm{F}}}$. \textbf{b,} Angle ($\vec{E}$ vs. $\vec{B}$) dependence of the chiral coefficient $C_{\textrm{W}}$. \textbf{c,} Density of states ($\nu(E)$) of the bulk electronic structure of TaAs shows a slow variation as a function of energy. The Berry curvature ($\Omega_x^2$) increases dramatically at the energy close to the Weyl nodes. \textbf{d,} Distribution of the square of the Berry curvature as a function of $k_z$ and energy $E$, evidencing that the Weyl points are the dominant source of Berry curvature. The plot is integrated with respect to $kx$ and $ky$ over the whole Brillouin zone. \textbf{e,}Temperature dependence of the axial charge relaxation time $\tau_{\textrm{a}}$ for sample a1. \textbf{f,} A cartoon illustrating the chiral anomaly based on our LMR data. The chiral anomaly leads to the axial charge pumping, $\vec{E}\cdot\vec{B}$. This causes a population imbalance difference between the Weyl cones with the opposite chiralities. The charge pumping effect is balanced by the axial charge relaxation, characterized by the time scale $\tau_a$ \cite{Son13, Burkov14, Burkov_2015}. Note that the axial charge relaxation time $\tau_a$ can be directly obtained from the observed negative LMR data through the chiral coefficient $C_{\textrm{W}}=\frac{e^4\tau_{\textrm{a}}}{4\pi^4\hbar^4g(E_{\textrm{F}})}$. We also note that this is a cartoon that assumes the Fermi level at zero $B$ field is exactly at the Fermi level.}
\end{figure*}

\clearpage
\textbf{
\begin{center}
{\large \underline{Supplementary Information}: \\Signatures of the Adler-Bell-Jackiw chiral anomaly in a Weyl Fermion semimetal}
\end{center}
}

\clearpage
\begin{figure*}[h]
\centering
\begin{center}
\includegraphics[width=17cm]{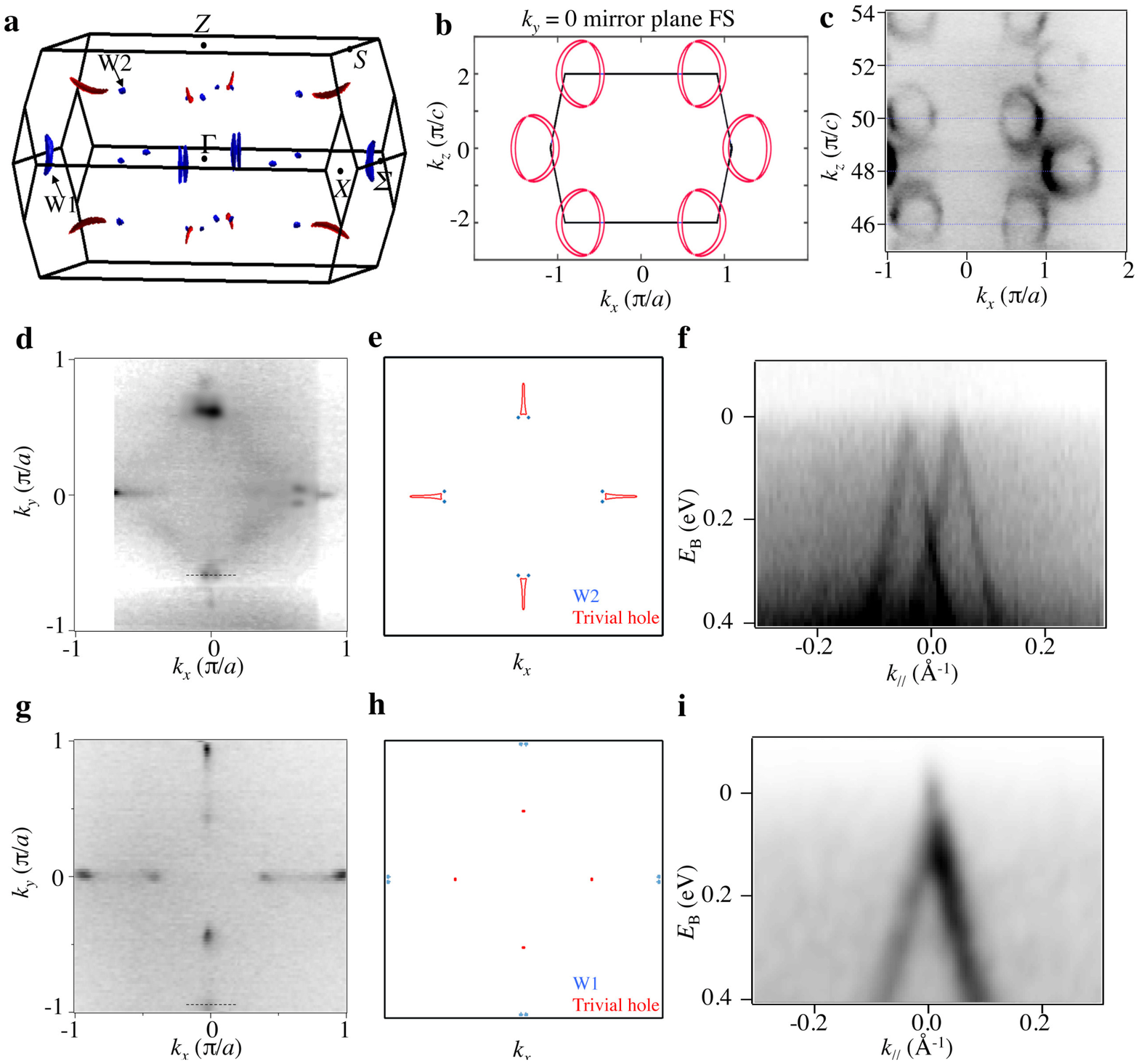}
\end{center}
\caption{\label{ARPES} \textbf{Systematic angle-resolved photoemission data.} \textbf{a,} First-principles calculated bulk Fermi surface in $k_x,k_y,k_z$ space. \textbf{b,c,} Calculated and ARPES measured bulk Fermi surface in $k_x,k_z$ space on the $k_y=0$ plane, showing the constant energy contours that arise from the trivial hole-like bands. \textbf{d,e,} ARPES measured and calculated bulk Fermi surface in $k_x,k_y$ space at the $k_z$ value that corresponds to the W2 Weyl nodes. \textbf{f,} ARPES $E-k_{//}$ dispersion map revealing the two nearby W2 Weyl cones. The direction of the dispersion cut is shown by the dotted line in panel (e). \textbf{g-i,} Same as panels (\textbf{d-f}) but for the W1 Weyl cones. }
\end{figure*}

\clearpage
\begin{figure*}[h]
\centering
\begin{center}
\includegraphics[width=17cm]{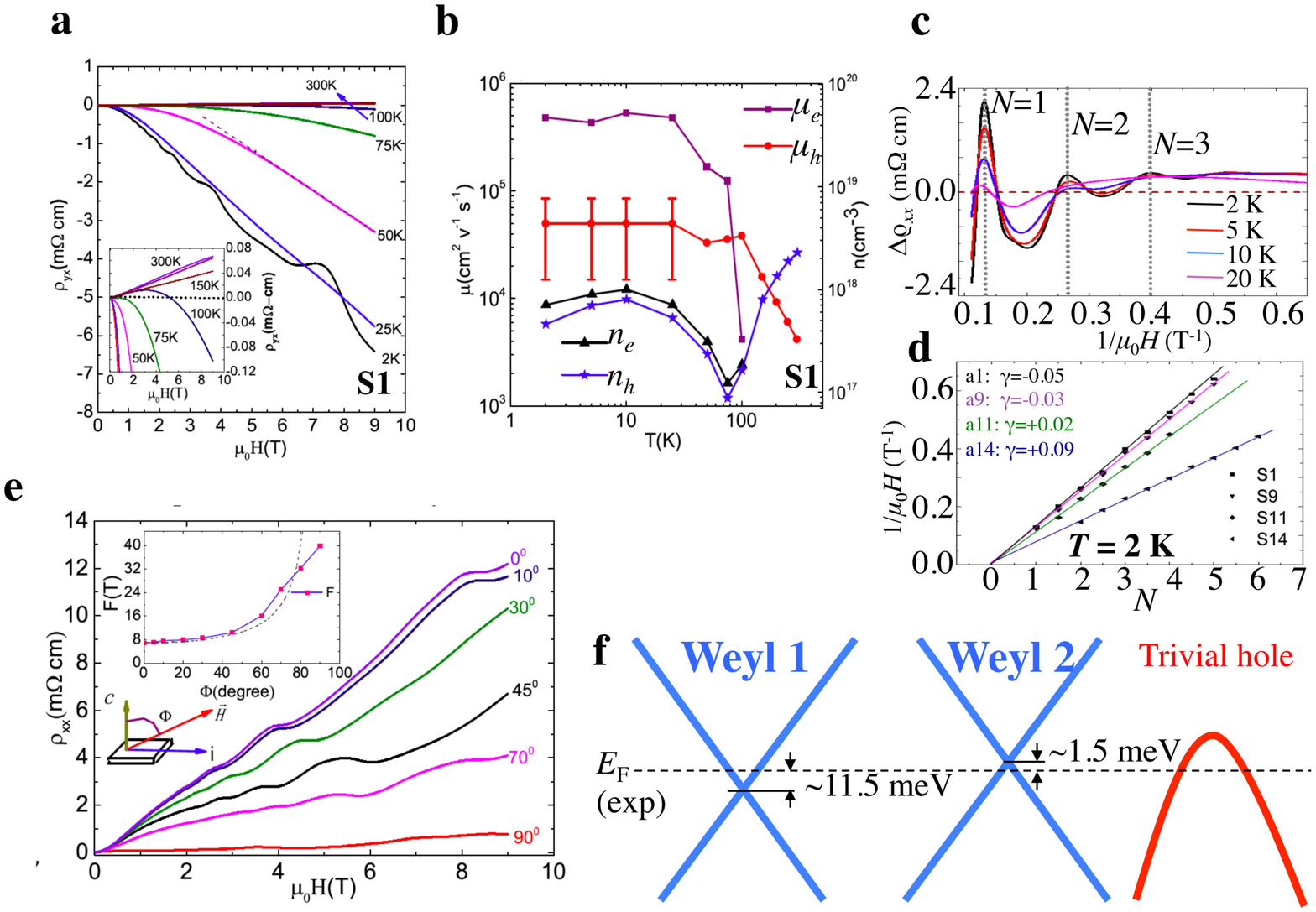}
\end{center}
\caption{\label{SdH} \textbf{Magneto-Transport and quantum oscillation data.} \textbf{a,} Hall resistivity versus the magnetic field in the temperature range from 2 to 300 K. Strong SdH oscillations were observed at 2 K. Inset: the high temperature Hall resistivity. \textbf{b,} Mobilities and carrier concentrations of the electrons and holes, clearly showing the coexistence of Weyl electrons and trivial holes in our samples. \textbf{c,} The oscillatory parts of $\sigma_{xx}$ at various temperatures, showing the $\pi$ Berry's phase of the Weyl electron pocket. \textbf{d,} The SdH fan diagram for four different samples. All of the four intercepts are located around zero, suggesting the $\pi$ Berry's phase of the Weyl electron pocket \textbf{e,} Magnetic field dependence of resistivity at representative $\Phi$ angles between $0^{\circ}$ - $90^{\circ}$ at 2K for sample a1, after heating the sample. The MR decreases rapidly when the magnetic field is tilted from \textit{c} to the direction of the current \textit{i}. Inset: the frequency F versus $\Phi$. The dashed curve is (1/$\cos\Phi$)$\cdot$F$_{0}$. \textbf{f,} Schematics of the experimentally determined band diagram indicating the positions of the two different types of Weyl nodes, W1 and W2, as well as the trivial hole-like band, relative to the experimentally determined Fermi level of sample a1.}
\end{figure*}

\clearpage
\begin{figure*}[h]
\centering
\begin{center}
\includegraphics[width=15cm]{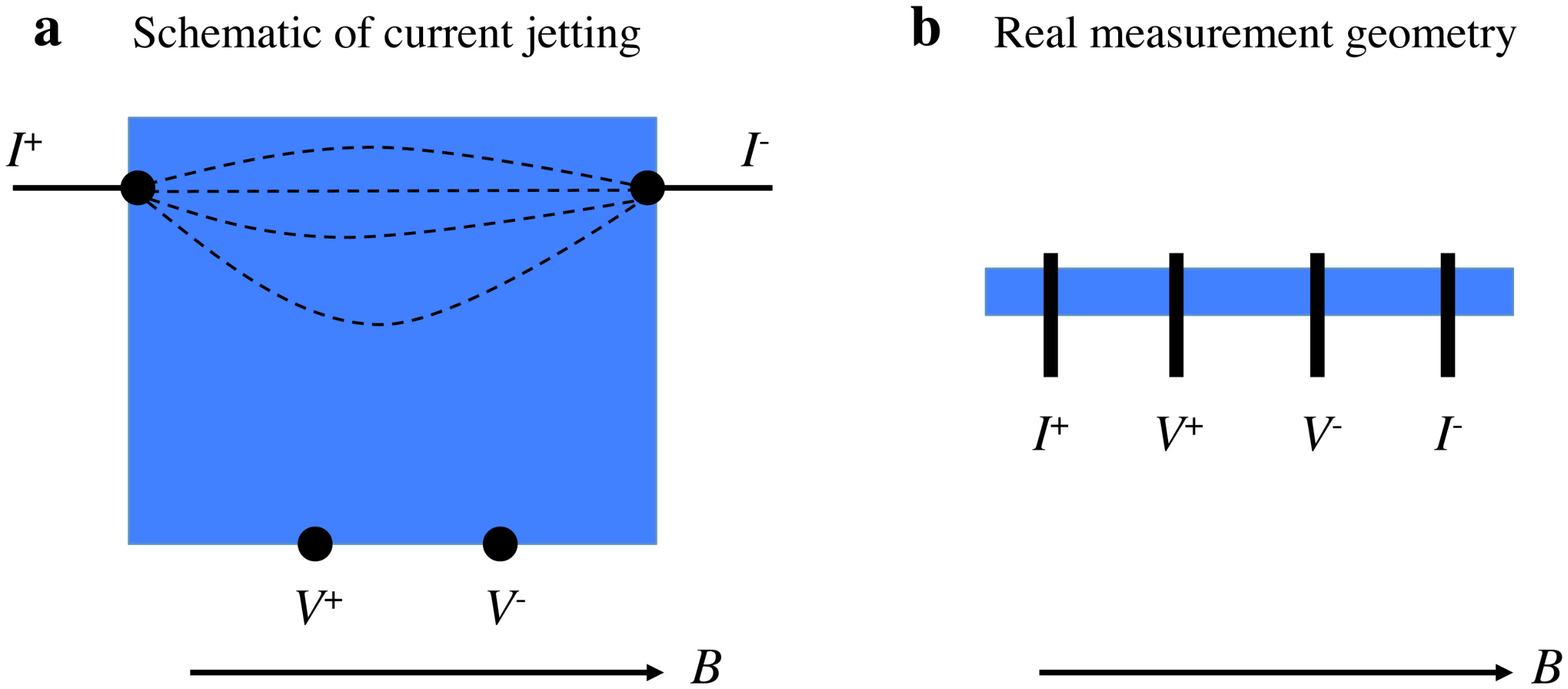}
\caption{\textbf{a,} A schematic illustration of the current jetting effect \cite{Pippard}. \textbf{b,} Our samples are carefully shaped to exclude the geometry/size effects.}\label{Jetting}
\end{center}
\end{figure*}

\clearpage
\begin{figure*}[h]
\centering
\begin{center}
\includegraphics[width=17cm]{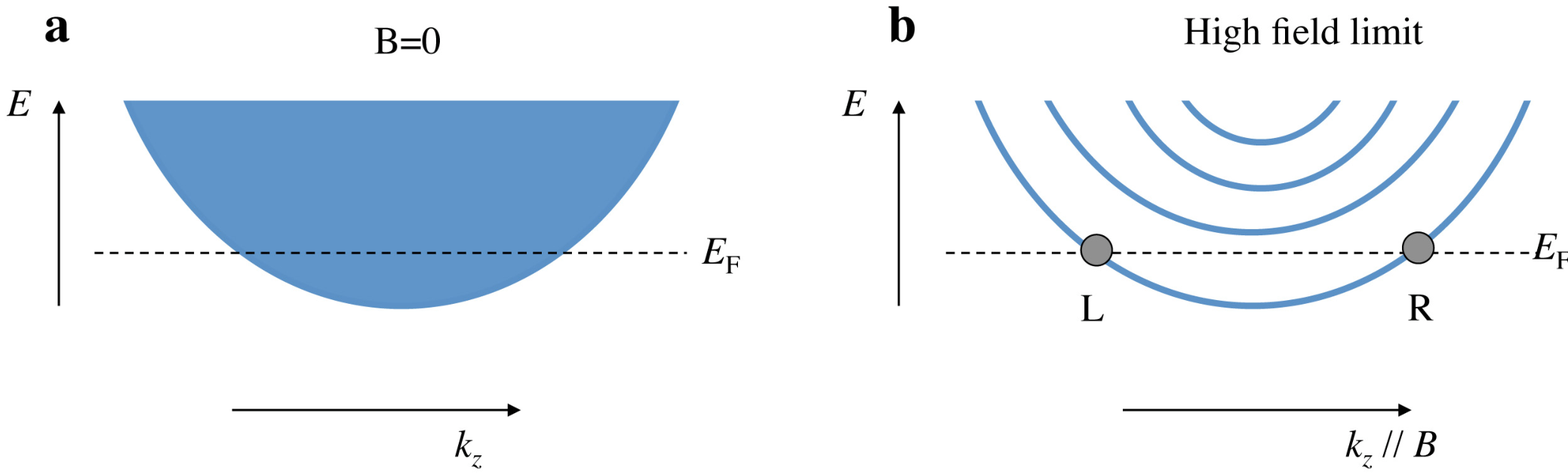}
\end{center}
\caption{\label{Quantum} \textbf{a}, Schematic band structure of a generic 3D metal without magnetic field. \textbf{b}, Schematic band diagram in the quantum limit regime.}
\end{figure*}

\clearpage
\begin{figure*}[h]
\centering
\begin{center}
\includegraphics[width=13cm]{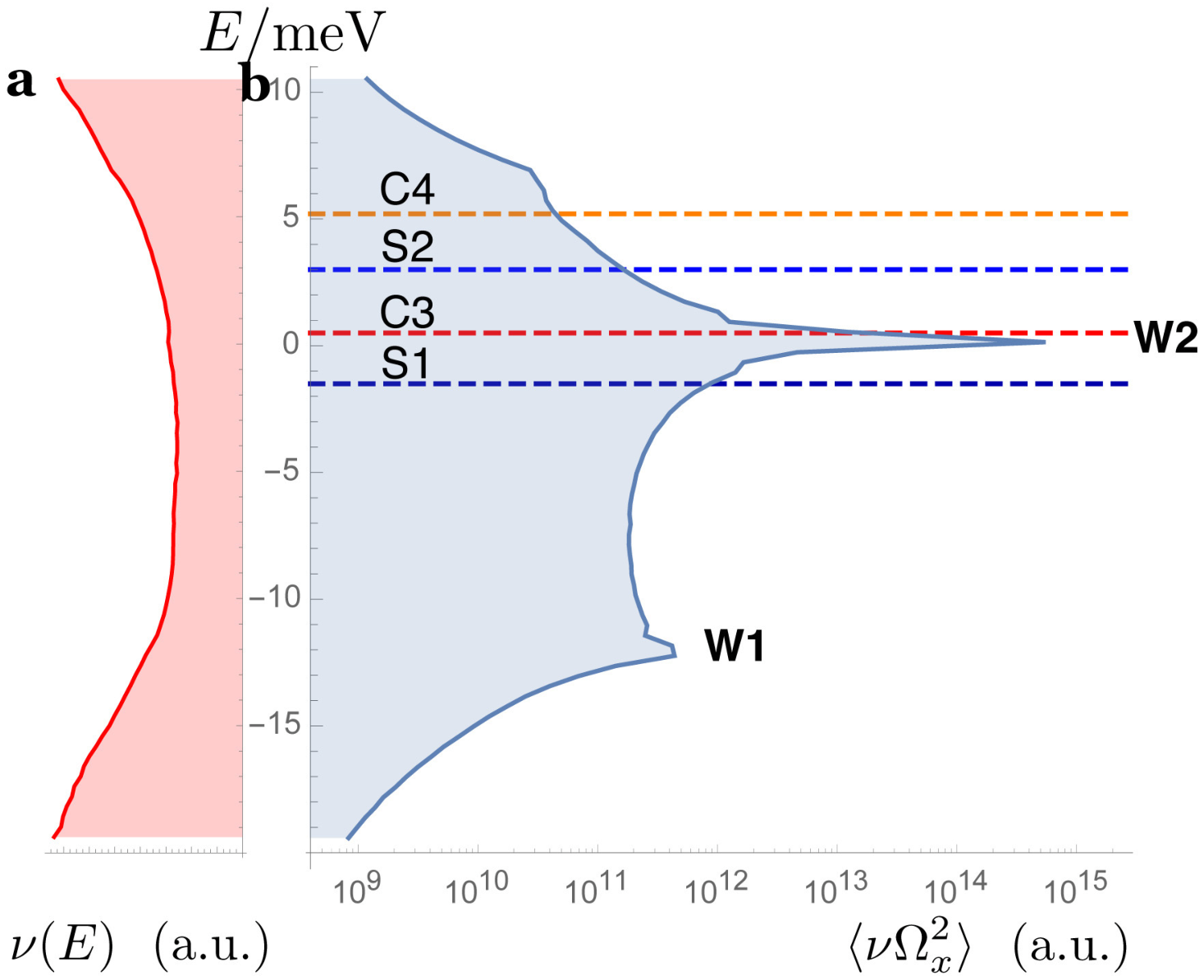}
\caption{\textbf{a,} Density of states of the bulk band structure of TaAs. \textbf{b,} Average of the square of the Berry curvature $\Omega_1$ over surfaces of equal energy in the bulk band structure of TaAs. The results were obtained using an effective $k\cdot{p}$ Hamiltonian that was fitted to the DFT band structure. While the density of states is featureless, the averaged Berry curvature is strongly enhanced near the Weyl nodes W1 and W2. This proves that the Weyl cones really dominate the contribution to the Berry curvature.}\label{fig: Berry}
\end{center}
\end{figure*}

\clearpage
\begin{figure*}[h]
\centering
\begin{center}
\includegraphics[width=10cm]{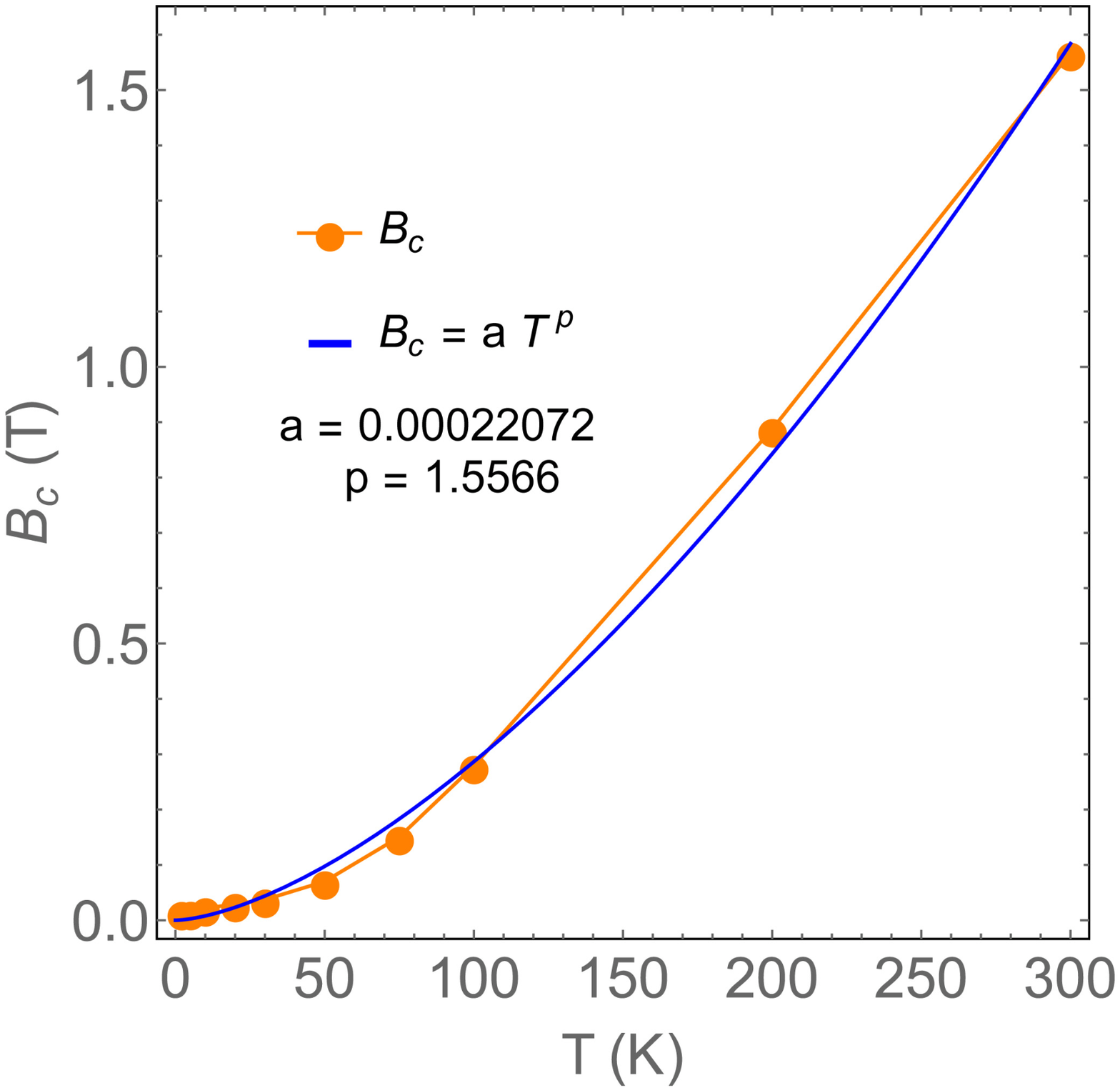}
\end{center}
\caption{\label{Bc} $B_c$ as a function of temperature obtained from fitting of the temperature dependent LMR data for sample a1 (Fig. 3a in the main text).}
\end{figure*}

\clearpage
\begin{table}
\begin{tabular}{|p{3cm}|p{3cm}|p{9cm}|}
\hline
DFT calculation & ARPES & Transport \\
\hline
$-8$ meV & $-5\pm1.5$ meV & Sample a1: $-1.5\pm0.86$ meV \\ \cline{3-3}
 & & Sample c2: $0.61\pm0.18$ meV\\ \cline{3-3}
& & Sample a3: $3\pm1.23$ meV\\ \cline{3-3}
& & Sample c4: $5.22\pm0.58$ meV\\ \cline{3-3}
& & Sample a5: $32.41\pm0.83$ meV\\
\hline
\end{tabular}

\caption{\label{EF} \textbf{Fermi energy ($E_{\textrm{F}}$) with respect to the W2 Weyl node.}}
\end{table}

\clearpage
\begin{center}
\begin{table}
\begin{tabular}{|p{5cm}|p{5cm}|p{5cm}|}
\hline
& $\omega\tau$ & LL index $N$ \\
\hline
Sample a1 & 1.2 & 115 \\
\hline
Sample c2 & 0.06 & 51 \\
\hline
Sample a3 & 0.3 & 30 \\
\hline
Sample c4 & 0.19 & 18 \\
\hline
\end{tabular}
\caption{\label{Hole} \textbf{Band parameters for the trivial hole pockets.} The $\omega\tau$ and Landau level index $N$ parameters of the trivial hole bands obtained from our transport measurements. These parameters are determined at the magnetic field of 0.3 T, which is a representative field value where the negative LMR takes place.}

\end{table}
\end{center}

\clearpage
\subsection{{\large Supplementary Note 1}}

In this section, we present systematic ARPES data to reveal the bulk electronic structure of TaAs. Our ARPES data show that there are three pockets that cross the Fermi level, namely, W1 Weyl cones, W2 Weyl cones and the trivial hole pockets. Supplementary Figure~\ref{ARPES}\textbf{c} shows the ARPES Fermi surface on the $k_y=0$ mirror plane. We observe ring-shaped hole-like contours. These contours demonstrate the trivial hole pockets, consistent with the calculations (Supplementary Figure~\ref{ARPES}\textbf{c}). In Supplementary Figure~\ref{ARPES}\textbf{d}, we show ARPES measured $k_x,k_y$ Fermi surface at the $k_z$ value that corresponds to the W2 Weyl nodes. We observe that the Fermi surface consists of discrete points, which are the W2 Weyl nodes. The energy dispersion measurement in Supplementary Figure~\ref{ARPES}\textbf{f} indeed reveals two cones, which are the two nearby W2 Weyl cones. Similarly, Figs.~\ref{ARPES}\textbf{g-i} are data showing the W1 Weyl cones and Weyl nodes. We find that the energy of the W1 Weyl node is about 10-20 meV lower than that of the W2 Weyl node (Figs.~\ref{ARPES}\textbf{f,i}), in agreement with the calculation (13 meV). More precisely determining their energy difference is limited by the experimental energy resolution ($>40$ meV). Because the $k$-space splitting of the W1 Weyl nodes is much smaller than that of the W2, resolving the two nearby W1 Weyl cones is beyond the experimental resolution. We emphasize that our ARPES data are important in the following aspects. First, our ARPES data experimentally prove the existence of Weyl fermions in our TaAs samples. Second, they experimentally reveal that there are three kinds of pockets crossing the Fermi level, W1 Weyl cones, W2 Weyl cones and the trivial hole bands. Third, they experimentally determine the Fermi energy, which is quite close to the W2 Weyl nodes. Thus W1 Weyl cones contribute electron pockets and the trivial bands contribute hole pockets.The W2 Weyl cones have small carrier density, and can be either electron-like or hole-like depending on the specific position of the Fermi energy with respect to the W2 nodes. These observations are crucial for our transport measurements.

\bigskip
\subsection{{\large Supplementary Note 2}}

We carried out magneto-transport and quantum oscillation measurements to determine the band parameters of the TaAs samples in our transport experiments. In Figure~\ref{SdH}, we show the magneto-transport data on sample a1 as a representative example. Same measurements were also performed on other samples presented in the main text in order to obtain their band parameters. Figure~\ref{SdH}\textbf{a} shows the Hall resistance $\rho_{yx}$ of sample a1 at different temperatures, clearly revealing a coexistence of the electron and hole carriers. This is consistent with the ARPES and first principles results presented above. We obtain the carrier density $n$ and the mobility $\mu$ for the electron and hole carriers, shown in Supplementary Figure~\ref{SdH}\textbf{b}.

We analyze the SdH quantum oscillation data at $T=2$~K. We use the following expression to analyze the SdH oscillation data, $\rho_{xx}$ at $T=2$~K, for a 3D system \cite{murakawa2013detection}.

\begin{equation}\label{eq1}
\rho_{xx}={\rho_0}[1+A(B,T)\cos2\pi(F/B+\gamma)]
\end{equation}

\begin{equation}\label{eq2}
A(B,T)\propto exp(-2\pi^2k_BT_D/\hbar\omega_c)\frac{2\pi^2k_BT/\hbar\omega_c}{sinh(2\pi^2k_BT/\hbar\omega_c)},
\end{equation}

Here,  $\rho_0$ is the non-oscillatory part of the resistivity, $A(B,T)$ is the amplitude of the SdH oscillations, $B$ is the magnetic field, $\gamma$ is the Onsager phase, $T_D$ is the Dingle temperature, $k_B$ is the Boltzmann's constant, $\omega_c$ is the cyclotron frequency, and $F=\frac{\hbar}{2\pi{e}}A_{\mathrm{F}}$ is the frequency of the oscillations. $A_{\mathrm{F}}$ is the extremal cross-sectional area of the Fermi surface (FS) associated with the Landau level (LL) index $N$, $e$ is the electron charge, and $2\pi\hbar$ is the Planck's constant. For sample a1, we obtain a Fermi surface area of $A_{\mathrm{F}}=7.07\times10^{-4}\ \mathrm{\AA}^{-2}$ and a Fermi wave vector $k_{\mathrm{F}}$ is $\sqrt{A_{\mathrm{F}}/\pi} = 0.015\ \mathrm{\AA}^{-1}$. We note that since the magnetic field is parallel to the $c$ crystallographic axis, the obtained Fermi surface area corresponds to the 2D cross-section of the 3D Fermi pocket that is perpendicular to the $k_z$ direction.

The Landau level index $N$ is plotted as a function of the inverse of the magnetic field strength ($1/\mu_0H$) in Supplementary Figure~\ref{SdH}\textbf{d}, from which one can see that, for all four samples, the linear interpolation of the curve intersects with the $x$ near zero ($\gamma=0$). This suggests that the electron carriers arise from a linearly dispersive band with a non-trivial Berry's phase \cite{murakawa2013detection}, consistent with the ARPES and first principles results that the electron carriers mainly arise from the W1 Weyl cones.

In order to obtain the Fermi velocity, the energy position of the chemical potential, and other important band parameters, we apply \ref{eq2}. We obtained a cyclotron mass $m_{\mathrm{cyc}}$ of $0.15m_{\mathrm{e}}$. We obtain the Fermi wave vector $k_{\mathrm{F}}$ is $\sqrt{A_{\mathrm{F}}/\pi} = 0.015\ \mathrm{\AA}^{-1}$, and the Fermi velocity $v_F$  is ${\hbar}k_{\mathrm{F}}/m_{\mathrm{cyc}} = 1.16\times10^5$~$\mathrm{m/s}$. Using a linear dispersion of this electron pocket, we obtain the chemical potential (relative to the energy of the W1 Weyl node) to be $E_{\mathrm{F}}$ = $m_{\mathrm{cyc}}v_{\mathrm{F}}^2 = 11.48$ meV for sample a1. Note that different samples have slightly different values for the chemical potential and other band parameters. This variation helps us to study the systematic dependence of the negative LMR, which is crucial for understanding its origin.

To further confirm that the observed electron carriers indeed arise from the W1 Weyl cones, we study the anisotropy of the electron-like pocket by tilting the magnetic field away from the $c$ direction in sample a1. Our data (Supplementary Figure~\ref{SdH}\textbf{e}) shows that the Fermi surface area along the $a$ axis is about 5 times larger that of along the $c$ axis. Therefore, the transport data show that the electron-like Fermi pocket is an ellipsoid that is elongated along the $c$ axis. We systematically check if the obtained band parameters from transport are consistent with ARPES and calculation. We place the chemical potential at 11.48~meV above the Weyl nodes W1 in our first-principles calculations and try to compare the calculated band parameters to those of obtained from transport. We have found an excellent agreement between calculation and transport: (1) The calculated carrier density of the electron pockets is $5.07\times10^{17}$~cm$^{-3}$, which agrees with our experimentally measured value in Supplementary Figure~\ref{SdH}\textbf{d}. (2) The anisotropy of the Fermi surface area is found to be 4.9, which is also in line with the experimentally determined value of 5.

Based on our systematic measurements, we obtain a band diagram presented in Supplementary Figure~\ref{SdH}\textbf{f}. The chemical potential lies $\sim11.5$~meV above W1. Therefore it is very close to W2, consistent with the ARPES results shown above.

Supplementary Table~\ref{Hole} shows the $\omega\tau$ ($\omega$ is the cyclotron frequency and $\tau$ is the transport lifetime) and the Landau level index $N$ parameters of the trivial hole bands determined from our transport measurements. These parameters define the (semiclassical or quantum) regime. Both $\omega\tau$ and $\nu$ are a function of the external magnetic field. The values in Supplementary Table~\ref{Hole} is obtained at the magnetic field of 0.3 T, which is a representative field value where the negative LMR takes place. Specifically, $\omega\tau$ is obtained by the following relationship,  $\omega\tau=\mu{H}$, where $\mu$ is the carrier mobility and $H$ is the magnetic field. The Landau level index $N$ is obtained from the Onsager relation (\ref{eq3}).
\begin{equation}\label{eq3}
 S_F(B)=\frac{2{\pi}eB}{\hbar}(N+\gamma),
\end{equation}
It can be seen that for the magnetic field of 0.3 T, none of the samples satisfy $\omega\tau>>1$ and Landau level index $N=0$ ($E_{\textrm{F}}$ only crosses the lowest LL). Thus the hole bands are not in the quantum limit. This is reasonable because the negative LMR is observed at quite low field (e.g. 0.3 T).

In order to extract the information of the carriers, the Hall conductivity tensor  $\mathrm{\sigma_{xy}=\rho_{yx}/(\rho_{xx}^2+\rho_{yx}^2)}$ was fitted by adopting a two-band model derived from the two-band theory \cite{Colin},
\begin{equation}
\sigma_{xy}=[n_h\mu_h^2\frac{1}{1+(\mu_hH)^2}-n_e\mu_e^2\frac{1}{1+(\mu_eH)^2}]eH,
\end{equation}
Where $n_e$ ($n_h$) and $\mu_e$ ($\mu_h$) denote the carrier concentrations and mobilities for the electrons (holes), respectively. Since there are four free parameters in this formula, we applied two constraints in the fitting process \cite{Ando}. The first constraint is the zero field resistivity.The second constraint we adopted is the Hall resistivity in the large B-field limit. In high field, the Hall resistivity reads $\rho_xy=1/{ec}\times1/(n_e-n_h)\times{B}$, so we can find the value of $n_e-n_h$ by a linear fitting of the high field data. These two constraints are standard constraints in transport works, as reviewed in Ref. \cite{Ando}. Therefore, the total number of free parameters in fits was be reduced to two.

\bigskip
\subsection{{\large Supplementary Note 3}}

In the Supplementary Table~\ref{EF}, we provide the Fermi energy of the TaAs samples determined by different approaches.

The ``$+$'' and ``$-$'' sign in the table means the $E_{\textrm{F}}$ being above or below the energy of the W2 Weyl node, respectively.The DFT calculated Fermi level is the norminal Fermi level directed obtained from the DFT calculation. The ARPES Fermi level was obtained by a linear fitting of the low energy Weyl fermion dispersion data (e.g. Fig. 1a). The transport Fermi level data was obtained from the SdH oscillation data. We note that the Fermi energy of the sample a5 is quite different from the others because sample a5 was grown by a different method where a different agent was used in the chemical vapor transport growth process. The error bars in this table are the errors in the fitting process.

\bigskip
\subsection{{\large Supplementary Note 4}}

We provide a diagrammatical highlight of the logical sequence for the experiments.

\textbf{\underline{Step-0} Basic spirit:} Magnetoresistance is a quite complicated phenomenon \cite{Spin1, Spin2, Spin3, Spin4, Spin5, Spin6, Pippard, Ag2Se, Argyres, Kikugawa, Das_Sarma, Haizhou, InSb, Burkov_2015}. A negative LMR by itself, although quite rare, is not a unique signature of the Weyl fermions \cite{Spin1, Spin2, Spin3, Spin4, Spin5, Spin6, Pippard, Ag2Se, Argyres, Kikugawa, Das_Sarma, Haizhou, InSb, Burkov_2015}.

In all previous studies \cite{Korean_BiSb, IOP_Chiral, Shuang_Chiral, Ong_Chiral,CdAs_Chiral, Yan_Chiral} reporting the chiral anomaly in Weyl and Dirac semimetals, the authors started by assuming that the negative LMR arises from the chiral anomaly. However, the key aspect that was missed is that it is entirely possible that these systematic dependence can be also consistent with other origins of the negative LMR. Specifically, none of the works have listed all possible origins for a negative LMR, and none of the works have discussed how one can distinguish each of the other origins from the chiral anomaly.

This is what has been achieved in this paper. We have considered comprehensively the possible origins for a negative LMR, and we have presented systematic data and analyses that lead us to the unique conclusion of the chiral anomaly due to Weyl fermions. In the process of doing that, we found that not only systematic LMR data (temperature, angular and other dependences) are necessary, but also having comprehensive information about the band structure at the Fermi level is crucial. Below we elaborate on the logical sequence of our work.

\bigskip
\textbf{\underline{Step-1} Negative LMR in magnetic element based materials:} The origin that is the easiest to exclude is the negative MR in magnetic materials, such as the giant magnetoresistance and the colossal magnetoresistance \cite{Spin1, Spin2, Spin3, Spin4, Spin5, Spin6}. These effects do not strongly depend on the angle between the electrical $\vec{E}$ and the magnetic $\vec{B}$ fields and occur in both the transverse MR and longitudinal MR. This is not consistent with our observations because our TaAs sample is nonmagnetic and because the MR in our experiment is only negative in the presence of parallel electrical $\vec{E}$ and magnetic $\vec{B}$ fields.

\bigskip
\textbf{\underline{Step-2} Classical (extrinsic) geometry or size effects:} A negative LMR can arise from a number of classical (extrinsic) geometry or size effects. They have been explained in detail in Ref. \cite{Pippard}. Essentially, these effects are caused by the inhomogeneous spatial distribution of the current in the sample. For example, Supplementary Figure~\ref{Jetting}\textbf{a} shows a schematic of a scenario where the current jetting effect can take place. Because the current and voltage contacts are misaligned in a four-probe setting, the current is largely distorted to be localized in the upper part of the sample when the current $i$ is parallel to the magnetic field $H$ (as indicated by the dotted lines). This causes a decrease of the measured voltage, leading to a negative LMR \cite{Pippard}. Other geometry or size effects are quite similar. For example, in polycrystalline samples, the inhomogeneity of the sample can also distort the current leading to a negative LMR. Similarly, anisotropy of the sample can cause similar effects.

In order to exclude the geometry and size effects, the samples were shaped into long, thin bars (thickness $<100$ $\mu$m) with four silver paste contacts fully crossing their width (Supplementary Figure~\ref{Jetting}\textbf{b}). This precaution can effectively prevent the current from distorting. Moreover, we measured many samples with different sizes (thickness). All samples with different thickness exhibited comparable negative LMR. This is inconsistent with the size or geometry effect, which is expected to depend strongly on the size and the shape of the samples. Furthermore, we have shown that the observed negative LMR is completely suppressed at temperatures above 50 K. By contrast, the negative LMR from geometry or size effects were found to survive even at room temperatures \cite{Ag2Se}. We also note that we observe the negative LMR with current flowing both along the crystallographic $a$ and $c$ axes. We note that TaAs is a tetragonal lattice. Hence the $a$ and $c$ axes represent the largest anisotropy that the system can offer. The fact that the negative LMR is observed along both $a$ and $c$ axes proves that it is irrelevant to the anisotropy of the system.

\bigskip
\textbf{\underline{Step-3} Chiral Landau levels in the quantum limit:} It has been theoretically pointed out \cite{Argyres, Das_Sarma} that a negative LMR can occur in a general 3D metal if the system is in the quantum limit regime, meaning that (1) one has $\omega\tau>>1$ ($\omega$ is the cyclotron frequency and $\tau$ is the transport lifetime) and that (2) the chemical potential only crosses the lowest Landau level. This is quite intuitive because in the ultra quantum limit, the system essentially has a pair of 1D chiral fermions with the opposite chiralities that arise from the lowest Landau level (Supplementary Figure~\ref{Quantum}\textbf{b}). Apparently, this construction \cite{Argyres, Das_Sarma} does not depend on the band structure details and can be realized in a generic 3D metal in the ultra quantum limit.

In order to check whether our negative LMR is due to this mechanism, we have carefully studied in which regime (semiclassical or quantum) our samples are located at the magnetic fields corresponding to the negative LMR. We note that the negative LMR are observed in low fields (e.g. $0.1$ T $\leq{B}\leq0.5$ T for sample a1). We have checked it quantitatively (see Table \ref{Hole}). For all samples studied, the system is always in the semiclassical limit.

We further emphasize the tricky nature of the LMR in the quantum (large $B$ field) limit. First, as mentioned above, Ref. \cite{Das_Sarma} showed that it is possible for a negative LMR to occur in a general 3D metal in the quantum limit. Second, Refs. \cite{Das_Sarma, Haizhou} showed that the sign of the LMR in fact depends on the nature of the impurities in the quantum limit \cite{Das_Sarma, Haizhou}. In fact, it is even theoretically shown that the Weyl cones that respect time-reversal symmetry can contribute a positive (not a negative) LMR in the quantum limit if the field dependence of the scattering time and Fermi velocity of the Landau bands is fully respected \cite{Haizhou}. Therefore, the sign of the LMR in the quantum regime seems to depend on the details of the impurities and scattering mechanisms. It cannot provide strong evidence for the Weyl fermions or the chiral anomaly.

\bigskip
\textbf{\underline{Step-4} Berry curvature in the semiclassical limit:} In the semiclassical limit, a negative LMR can occur if the band structure has a nonzero Berry curvature at the Fermi level \cite{Burkov_2015}. The negative LMR due to Weyl fermions belongs to this case because Weyl nodes serve as sources or drains of Berry curvature. However, it is also important to note that theoretically any band with a nonzero Berry curvature will contribute a term in the equation of motion that leads to a negative LMR \cite{Burkov_2015}. This means that, even up to here, after we have excluded many origins above, we can show that the observed LMR is due to the nonzero Berry curvature of the band structure in the semiclassical limit. We cannot yet show that it is uniquely due to the Weyl fermions, let alone the chiral anomaly.

Thus we further show two pieces of evidence to establish the connection between the negative LMR and the Weyl fermions in TaAs. (1) We have studied the contribution of Berry curvature from each bands carefully (see Supplementary Figure~\ref{fig: Berry}). We show that in our TaAs system the Berry curvature almost entirely arises from the Weyl cones. The contribution from the trivial hole like bands is negligible. (2) We have fitted the negative LMR and found that the chiral coefficient $C_{\textrm{W}}$ has a $\frac{1}{E_{\textrm{F}}^2}$ dependence. Note that the $\frac{1}{E_{\textrm{F}}^2}$ dependence of the chiral coefficient $C_{\textrm{W}}$ is a result of the linear dispersion and the specific Berry curvature distribution of the Weyl cones. In order words, if one started with a different band (not Weyl) that has a different dispersion and a different distribution of the Berry curvature, then the expression of the chiral coefficient would have been different and the $C_{\textrm{W}}\propto\frac{1}{E_{\textrm{F}}^2}$ dependence would have been invalid. These data and analyses show that the observed negative LMR is not due to the trivial hole bands. Instead, it arises from the Weyl fermions in TaAs.

We emphasize that this last step is crucial. One of the reasons that this can be achieved in our study is that we have systematically mapped out the band structure using three independent ways (first-principles, ARPES and quantum oscillations). All the other systematic dependences, including the ($\vec{E}$ vs $\vec{B}$) angle, temperature, and the current direction with respect to the crystallographic axis, which are presented here and also in Ref. \cite{Korean_BiSb, IOP_Chiral, Shuang_Chiral, Ong_Chiral,CdAs_Chiral, Yan_Chiral}, can \textit{\textbf{not}} distinguish the negative LMR due to Weyl fermions from the negative LMR due to other band structures with nonzero Berry curvature. In other words, if a negative LMR were induced by a (non-Weyl) band structure due to its nonzero Berry curvature, then it will also show quantitatively the same angular, temperature, and current direction dependences. Therefore, we emphasize that it is crucial to have full information of the band structure of the system studied in transport. Also, the $\frac{1}{E_{\textrm{F}}^2}$ dependence of the chiral coefficient $C_{\textrm{W}}$, which is uniquely presented here not in other studies, is crucial because it really depends on the details of the band dispersion and Berry curvature distribution of the Weyl cones, not just the fact that the bands have some nonzero Berry curvature.

\bigskip
\subsection{{\large Supplementary Note 6}}

In order to calculate the Berry curvature of the band structure, we built a $k\cdot{p}$ model whose general structure was introduced in Ref. \cite{Weng2015}. In the absence of spin-orbit coupling, the Hamiltonian $H_0$ has a nodal ring in the $(k_2=0)$ plane that is protected by mirror symmetry.  When spin-orbit coupling is included, the degeneracy of the nodal ring is lifted by the mass terms $m_i,\ i=1,\cdots, 6$ that contribute to $H_{\mathrm{mass}}$ below. The combinations of these mass terms can give rise to pairs of Weyl nodes off the $(k_2=0)$ plane.

A pair of Weyl nodes in the $(k_3=0)$-plane are generated by the combination of the masses $m_4$ and $m_6$, while the Weyl nodes away from the $(k_3=0)$-plane stem from the interplay of $m_4$ and $m_5$.

The total Hamiltonian is given by
\begin{eqnarray}
H=&&H_0+H_{\mathrm{mass}},\\
H_0=&& \epsilon(\mathbf{k})\sigma_0+d_1(\mathbf{k})\sigma_1+d_2(\mathbf{k})\sigma_2+d_3(\mathbf{k})\sigma_3,\\
H_{\mathrm{mass}}=&&
m_1(\mathbf{k})\sigma_0s_2+m_2(\mathbf{k})\sigma_3s_2+m_3(\mathbf{k})\sigma_1s_1+\,m_4(\mathbf{k})\sigma_1s_3+m_5(\mathbf{k})\sigma_2s_3
\nonumber\\
&&+m_6(\mathbf{k})\sigma_2s_1,
\end{eqnarray}
where $s_\mu,\ \mu=0,\cdots, 3$, are the $2\times2$ identity matrix and the three Pauli matrices acting on the spin degree of freedom of the electrons and $\sigma_\mu,\ \mu=0,\cdots, 3$, are the $2\times2$ identity matrix and the three Pauli matrices acting on an effective orbital degree of freedom.

The $k_2\to -k_2$ mirror symmetry, as well as the combination of time-reversal and $C_2$ rotation symmetry, impose whether each term is even or odd under $k_2\to -k_2$ and under $k_3\to -k_3$. We expand each term up to a given order in $\mathbf{k}$, including only symmetry-allowed contributions:


\begin{eqnarray}
\varepsilon(\mathbf{k})&=&\mu+w k_1 +\mathcal{O}(\mathbf{k}^2),\\
d_1(\mathbf{k})&=&uk_2k_3+\mathcal{O}(\mathbf{k}^3),\\
d_2(\mathbf{k})&=&vk_2+\mathcal{O}(\mathbf{k}^2),\\
d_3(\mathbf{k})&=&M-ak_1^2-bk_3^2+c k_1
\nonumber\\
&&+d k_1^3+\mathcal{O}(\mathbf{k}^3),\\
m_1(\mathbf{k})&=&m_1 +\mathcal{O}(\mathbf{k}),\\
m_2(\mathbf{k})&=&m_2 +\mathcal{O}(\mathbf{k}),\\
m_3(\mathbf{k})&=&m_3 k_3+\mathcal{O}(\mathbf{k}^2),\\
m_4(\mathbf{k})&=&m_4 +m_4^\prime k_1 +\mathcal{O}(\mathbf{k}^2),\\
m_5(\mathbf{k})&=&m_5 k_3+\mathcal{O}(\mathbf{k}^2),\\
m_6(\mathbf{k})&=&m_6 +\mathcal{O}(\mathbf{k}).
\end{eqnarray}

We then fitted this $k\cdot{p}$ Hamiltonian to the first-principle band structure with the goal to reproduce the ring-shaped trivial Fermi surface and the correct location and number of Weyl nodes. This is achieved with the parameters $M=12.23$, $\mu=-3.504$, $u=-763.1$, $v=-685.1$, $w=34.11$, $a=682.8$, $b=583.0$, $c= 264.2$, $d= -147.5$, $m_1=7.019$, $m_2=1.031$, $m_3=0.9078$, $m_4=0.0$, $m_4^\prime=-11.07$, $m_5=-56.50$, $m_6=-4.097$, all in units of meV and the appropriate power of $\textrm{\AA}$.

This effective Hamiltonian was used to compute the Berry curvature $\Omega_{i}=\mathrm{i}\epsilon_{ijl}\langle \partial_{k_j}u(\mathbf{k})|\partial_{k_l}u(\mathbf{k})\rangle$, where $|u(\mathbf{k})\rangle$ are nondegenerate Bloch states. (For the isolated degeneracy points in the BZ we have to consider the non-Abelian form of the Berry curvature.)

Figure~\ref{fig: Berry} shows the average of $\Omega_1^2$ over a contour of equal energy $E$ weighted with the density of states $\nu(E)$ at this energy. While the density of states is featureless, the averaged Berry curvature is strongly enhanced near the Weyl nodes W1 and W2. We clearly observe two peaks in $\langle\nu\Omega_1^2\rangle$ which stem from the dominant Berry curvature near the Weyl nodes W1 and W2 (notice the log scale). Figure~\ref{fig: Berry} also shows that no other anomalous sources of diverging Berry curvature exist in the band structure, besides the Weyl nodes. (The actual divergence at the Weyl points is cut off due to the numerical accuracy.) This proves that the Weyl cones really dominate the contribution to the Berry curvature, and that the contribution from the trivial hole like bands is negligible.

\bigskip
\subsection{{\large Supplementary Note 6}}

As described in the main text, we use the following equation to fit the LMR data.

 \begin{equation}\label{Fit}
\sigma_{xx}(B)=8{C_{\textrm{W}}}B^2-C_{\textrm{WAL}}\bigg(\sqrt{B}\frac{B^2}{B^2+B_{\textrm{c}}^2}+\gamma{B}^2\frac{B_{\textrm{c}}^2}{B^2+B_{\textrm{c}}^2}\bigg)+\sigma_0
\end{equation}

Here we discuss additional details regarding the fitting:

Firstly, we discuss the 3D weak anti-localization (WAL) term. The 3D WAL effect, the $C_{\textrm{WAL}}$ term in the fitting formula, accounts for the initial steep uprise of the LMR at small magnetic fields. In the fitting formula, we have included a critical field $B_{\textrm{c}}$ that characterizes the crossover from a $-B^2$ dependence near zero field to $-\sqrt{B}$ dependence at higher fields. Here we describe the underlying physical reason for this crossover carefully. $B_c$ is related to the phase coherence length $\ell_\phi$. At low temperatures and no intervalley scattering, $\ell_\phi\rightarrow \infty$, theory predicts a $-\sqrt{B}$ dependence of the WAL \cite{Haizhou}. This means that one has $B_c\simeq0$. In the other limit, meaning high temperatures or in the presence of strong inter-valley scattering, $\ell_\phi\rightarrow 0$, and a $-B^2$ dependence is theoretically expected \cite{Haizhou}. This means that $B_c$ is large. Roughly,
\begin{eqnarray}
B_c\sim \frac{\hbar}{e\ell_\phi^2}.
\end{eqnarray}
Empirically, the temperature dependence of the phase coherence length can be written as $\ell_\phi \sim T^{-p/2}$, then $B_c \sim (\hbar/e) T^p$, where $p$ is positive and determined by decoherence mechanisms such as electron-electron interaction ($p=3/2$) or electron-phonon interaction ($p=3$). Also, We expect that the intervalley scattering may correct $\ell_\phi$ from these simple power law.

In Fig. 3a of the main text, we have presented the temperature dependent LMR data for sample a1 and their fits. Here in Supplementary Figure~\ref{Bc} we show the corresponding $B_c$ as a function of temperature. It can be seen that at low temperatures the $B_c$ is almost zero, which indicates that the WAL follows well the $-\sqrt{B}$ dependence. At higher temperature, $B_c$ increases monotonically. We fit the $B_c$ as a function of $T$, from which we extract that $p\approx 1.5$, indicating that the electron-electron interaction is the dominant decoherence mechanism.

Secondly, regarding the  $\sigma_0$ term in the fitting formula (\ref{Fit}). We note that it contributes the positive LMR that arises from the Drude conductivity of conventional charge carriers present in TaAs. In parallel fields, the Lorentz force is zero so the Drude conductivity is a constant, under the assumption that the corresponding Fermi surface is isotropic. In the case of TaAs, the Fermi surface that gives rise to the positive LMR is the trivial hole pocket, which is quite anisotropic. Thus a weak magnetic field dependence can also be possible. However, we note that even if there is a weak magnetic field dependence for the Drude conductivity in parallel fields, it has to be positive because the only source of negative LMR in the semiclassical limit is the Berry curvature. And as we have shown clearly in the main text, in TaAs, the Berry curvature comes from the Weyl cones, not the trivial hole pockets. In our case, we find that using a constant for the Drude conductivity already gives satisfactory results for the fitting of the LMR data, which is sufficient for our purpose. This also reduces free parameters in the fitting, making the fitting more robust.

Finally, we note that except the chiral $C_{\textrm{W}}$ term, all other terms in \ref{Fit} give rise to positive LMR. They are used to simulate the other effects (Drude conductivity or WAL) that coexist with the chiral charge pumping. Therefore, they are not the focus of our paper. The focus is the chiral $C_{\textrm{W}}$ term, which characterizes the chiral anomaly.


\bigskip
\subsection{{\large Supplementary Note 7}}
The $y$ axes of Figs. 2\textbf{a-e} and Figs. 3\textbf{a-b} are the change of the resistivity with respect to the zero-field resistivity, $\Delta\rho=\rho(B)-\rho(B=0)$. The zero-field resistivities are 5.65, 221.28, 7.02, 40.12, 12.59 ($\mu$Ohm cm) for Figs. 2\textbf{a-e}, respectively; 5.65, 6.10, 12.59, 10.32, 12.44, 19.91, 31.90, 46.54, 87.07, 122.53 ($\mu$Ohm cm) for temperature $T=$ 2, 5, 10, 20, 30, 50, 75, 100, 200, 300 K in Fig. 3\textbf{a}, and 5.43, 12.04, 7.56, 6.13, 7.91, 9.02, 6.82, 7.02, 8.02, 5.91, 8.27 ($\mu$Ohm cm) for angles 85, 75, 55, 40, 25, 15, 5, 0, -3, -5, -7 (degree) in Fig. 3\textbf{b}.

\vspace{2cm}
\*Correspondence and requests for materials should be addressed to
M.Z.H. (Email: mzhasan@princeton.edu) and S.J. (gwljiashuang@pku.edu.cn).

\end{document}